\documentclass[review, 3p]{elsarticle} 

\usepackage{lineno}
\usepackage[colorlinks, citecolor=blue]{hyperref}
\modulolinenumbers[5]

\usepackage{amsmath, amssymb}
\usepackage{multirow}
\usepackage{graphicx}
\usepackage{epstopdf,epsfig}

\usepackage{ntheorem}
\newtheorem{hyp}{Hypothesis} 
\usepackage{soul}
\usepackage{makecell}
\usepackage{tikz}
\usetikzlibrary{shapes}
\usepackage{xcolor}
\usepackage{dirtytalk}
\usepackage{float}
\usepackage{tabularx}
\usepackage{adjustbox}
\usepackage{enumitem}

\usepackage{tgpagella}
\usepackage{mathpazo}

\newcommand{\gaussone}{\texttt{Gauss1}~}
\newcommand{\gausstwo}{\texttt{Gauss2}~}
\newcommand{\roma}{\texttt{Roma}~}
\newcommand{\romagausstwo}{\texttt{Roma-Gauss2}~}

\journal{International Journal of Multiphase Flow}

\biboptions{authoryear}

\def\del{\partial}

\definecolor{myteal}{RGB}{40, 128, 128}
\definecolor{mycyan}{RGB}{0, 205, 205}
\definecolor{myblue}{RGB}{0, 0, 255}
\definecolor{mygreen}{RGB}{0, 128, 0}
\definecolor{mypurple}{RGB}{155, 48, 255}
\definecolor{myred}{RGB}{255, 0, 0}
\definecolor{mygold}{RGB}{238, 173, 14}
\newcommand{\bluecircle}{\raisebox{0.5pt}{\tikz{\node[draw=myblue,scale=0.4,circle,fill=myblue](){};}}}
\newcommand{\greentriangle}{\raisebox{0.5pt}{\tikz{\node[draw=mygreen,scale=0.3,regular polygon,regular polygon sides=3,fill=mygreen,rotate=0](){};}}}
\newcommand{\redtriangle}{\raisebox{0.5pt}{\tikz{\node[draw=myred,scale=0.3,regular polygon,regular polygon sides=3,fill=myred,rotate=0](){};}}}
\newcommand{\purplesquare}{\raisebox{0.5pt}{\tikz{\node[draw=mypurple,scale=0.4,regular polygon,regular polygon sides=4,fill=mypurple](){};}}}
\newcommand{\redx}{\raisebox{0pt}{\tikz{\node[draw=myred,scale=1,cross out,minimum size=4pt,inner sep=0pt,outer sep=0pt,line width=0.6mm,draw=myred](){};}}}
\newcommand{\greenx}{\raisebox{0pt}{\tikz{\node[draw=mygreen,scale=1,cross out,minimum size=4pt,inner sep=0pt,outer sep=0pt,line width=0.6mm,draw=mygreen](){};}}}
\newcommand{\golddiamond}{\raisebox{0pt}{\tikz{\node[draw=mygold,scale=0.4,diamond,fill=mygold](){};}}}

\newcommand{\reddowntriangle}{\raisebox{0pt}{\tikz{\node[draw=myred,scale=0.4,regular polygon,regular polygon sides=3,fill=myred,rotate=60](){};}}}
\newcommand{\greencircle}{\raisebox{0.5pt}{\tikz{\node[draw=mygreen,scale=0.4,circle,fill=mygreen](){};}}}
\newcommand{\purpletriangle}{\raisebox{0.5pt}{\tikz{\node[draw=mypurple,scale=0.3,regular polygon,regular polygon sides=3,fill=mypurple,rotate=0](){};}}}
\newcommand{\bluesquare}{\raisebox{0.5pt}{\tikz{\node[draw=myblue,scale=0.4,regular polygon,regular polygon sides=4,fill=myblue](){};}}}
\newcommand{\cyanx}{\raisebox{-1pt}{\tikz{\node[draw=mycyan,scale=1,cross out,minimum size=3pt,inner sep=0pt,outer sep=0pt,line width=0.6mm,rotate=45,draw=mycyan](){};}}}

\DeclareRobustCommand\tealdotted{\tikz[baseline=-0.6ex]\draw[myteal,very thick,dotted] (0,0)--(0.3,0);}
\DeclareRobustCommand\blackline{\tikz[baseline=-0.6ex]\draw[black,very thick,solid] (0,0)--(0.3,0);}

\begin{document}

\begin{frontmatter}

\title{Effect of interpolation kernels and grid refinement on two way--coupled point-particle simulations}

\author[label1]{Nathan A. Keane}
\address[label1]{College of Earth, Ocean, and Atmospheric Sciences, Oregon State University, Corvallis, OR 97331}

\author[label2]{Sourabh V. Apte}
\address[label2]{School of Mechanical, Industrial, and Manufacturing Engineering, Oregon State University, Corvallis, OR 97331}
\ead{Sourabh.Apte@oregonstate.edu, Corresponding Author}

\author[label3]{Suhas S. Jain}
\address[label3]{Center for Turbulence Research, Stanford University, Palo Alto, CA 94305}

\author[label3]{Makrand A. Khanwale}

\begin{abstract}
The predictive capability of two way--coupled point-particle Euler-Lagrange model in accurately capturing particle-flow interactions under grid refinement, wherein the particle size can be comparable to the grid size, 
 is systematically evaluated. Two situations are considered, (i) uniform flow over a stationary particle, and (ii) decaying isotropic turbulence laden with Kolmogorov-scale particles.
Particle-fluid interactions are modeled using only the standard drag law, typical of large density-ratio systems. A zonal, advection-diffusion-reaction (Zonal-ADR) model is used to obtain the undisturbed fluid velocity needed in the drag closure. 
Two main types of interpolation kernels, grid-based and particle size--based, are employed. The effect of interpolation kernels on capturing the particle-fluid interactions, kinetic energy, dissipation rate, and particle acceleration statistics are evaluated in detail.
 It is shown that the interpolation kernels whose width scales with the particle size perform significantly better under grid refinement than kernels whose width scales with the grid size. Convergence with respect to spatial resolution is obtained with the particle size--based kernels with and without correcting for the self-disturbance effect. While the use of particle size--based interpolation kernels provide spatial convergence and perform better than kernels that scale based on grid size, small differences can still be seen in the converged results with and without correcting for the particle self-disturbance. Such differences indicate the need for self-disturbance correction to obtain the best results, especially when the particles are larger than the grid size.\\

\end{abstract}


\end{frontmatter}



\section{Introduction}
\label{sec: intro}
Many engineering, biological, and environmental applications involve disperse particle-laden flows, wherein small size solid particles, liquid droplets, or gaseous bubbles are dispersed in a fluid flow, such as sediment transport, fluidized beds, spray injectors in gas-turbine combustion chambers, cavitation, among others. When the number of dispersed particles is very large, on the order of ${\mathcal O}(10^6)$--${\mathcal O}(10^9)$, 
the point-particle (PP) approach~\citep{maxey1983} is commonly employed owing to its simplicity and affordability.

In standard point-particle models, the particles are assumed to be spherical, significantly smaller than the grid size or the length scale of the smallest resolved flow features, have low volume loading, and be modeled as point sources of mass, momentum, and energy. Particle dynamics are modeled by solving the Maxey-Riley equations with force closures for drag, lift, added mass, pressure, history forces, among others~\citep{maxey1987}. In order to couple the particle and fluid phases in a two-way coupled framework, the fluid velocity interpolated at the particle location is used in the force closure models, and a reaction force from the particles is added to the fluid momentum equation (the energy interactions are also represented in a similar way for compressible flows with heat transfer). 
Many important studies have used this point-particle model to investigate particle-turbulence interactions \citep{squires1990, elghobashi1991, elghobashi1993, Boivin.Simonin.ea1998, Ferrante.Elghobashi2003}. 
The particle force closures are typically based on the relative slip velocity at the particle location, which involves the difference between {\it the undisturbed fluid velocity} seen by the particle and the particle velocity, especially for low-volume loadings. The undisturbed fluid velocity seen by the $p^{th}$ particle is defined as what the flow velocity would be in the absence of the $p^{th}$ particle, but with all other particles present. The undisturbed fluid velocity associated with each particle is not readily available. It is a standard practice to simply use the two way-coupled fluid velocity for computing particle force closures~\citep{apte2003large}.

When particles are very small compared to the grid size ($D_p/\Delta \ll 1$), where $D_p$ is the particle diameter and $\Delta$ is the grid size, the above approximation (using two-way coupled fluid velocity) does not result in  significant error, as this is closer to the original assumptions of a point-particle model. However, when the particle size becomes comparable to the grid size ($D_p/\Delta \sim 1$), using the two way--coupled disturbed flow field in the force closure models can lead to significant errors in particle and fluid statistics, especially at low particle Reynolds numbers. ~\citet{burton2005fully} conducted particle-resolved direct numerical simulations of a single, fixed particle of size $D_p \approx 2\eta$ in decaying isotropic turbulence, where $\eta$ is the Kolmogorov length scale. They found that the instantaneous error in modeled particle force varied between 15-30\% with standard point-particle model without correction for the particle self-disturbance.~\citet{hwang2006homogeneous} in a study on homogeneous, isotropic turbulence modulation by small, heavy particles concluded that the extra dissipation caused by particles of size comparable to the Kolmogorov scale was grossly underestimated by the point-particle model that do not correct for the self-disturbance created by the particle. In addition, these errors typically increase with decrease in grid size, resulting in different particle-fluid interactions under grid convergence~\citep{horwitz2020}. 
In a direct or large-eddy simulation of particle-laden flows, the particle size can be comparable to or even larger than the local grid resolution. For example, in wall-bounded flows fine grids are needed in the wall-normal direction for these computations. The particle size can be several times larger than the finest wall-normal grid resolution, and neglecting the effect of particle self-disturbance in closure models can lead to large errors. Similarly, in simulations of spray injectors, the droplet sizes can be much larger than the grid resolution near the injector~\citep{moin2006large}. 

Several recent studies have been devoted to quantifying the effect of self-disturbance when the particle becomes comparable to the grid size~\citep{gualtieri2015,horwitz2016,horwitz2018,esmaily2018,fukada2018numerical,liu2019self,pakseresht2020,pakseresht2021disturbance,horwitz2022discrete,balachandar2023correction,apteADR}. 
In addition to obtaining the undisturbed fluid velocity, the closure models for point-particle dynamics require an interpolation kernel that interpolates the fluid properties from the surrounding control volumes to the Lagrangian point-particle location, for example, to calculate the slip-velocity used in the force closures for the equations of particle motion. Similarly, in two-way coupled simulations, an interpolation kernel is used to distribute the particle forces back to the background Eulerian grid. Several different interpolation kernels have been used for the Eulerian-grid-to-Lagrangian-particle-location (E2L) and Lagrangian-particle-to-Eulerian-grid (L2E) interpolations. In general, these interpolation kernels can be classified into two categories based on the kernel width used: (i) grid-based kernels, wherein the kernel width is based on local grid size and is independent of the particle size, and (ii) particle size--based kernel, wherein the kernel width scales with the particle size. Typically the E2L and L2E interpolation functions are identical, following the kinetic energy conservation principles identified by~\citet{sundaram1996numerical}. However, majority of their analysis involved cases with particle size much smaller than the grid, $D_p/\Delta \ll 1$.

The grid-based interpolation kernels have been commonly used because of their ease of implementation on complex anisotropic or unstructured grids in three dimensions. These include trilinear~\citep{ferrante2005}, cubic splines~\citep{horwitz2020}, clipped Gaussian kernel~\citep{apte2003large,apte2009large}, among others. A Delta-function-based interpolation kernel with compact support~\citep{roma1999adaptive} has been widely used in immersed boundary-based methods. 
The main advantage of grid-based interpolations is that their compact support requires only the nearest neighbors of the control volume within which the particle center is located, making it attractive for complex grids. However, for anisotropic and unstructured grids, the grid-based interpolations can lead to asymmetric interpolation weights and potentially impact the accuracy of the simulations. However, these effects are not significant if the particle size is much smaller than the grid size.

Recently, \citet{horwitz2020} conducted a detailed evaluation of the numerical methods and the effect of grid size, relative to particle size on fluid-particle interactions predicted by point-particle models. They used grid-based interpolation kernels, with and without correcting for the particle self-disturbance field, in homogeneous isotropic turbulence. For a particle size to a Kolmogorov-scale ratio ($D_p/\eta$) of 0.25, the grid size ($D_p/\Delta$) was varied over the range of $0.25$--$1$, and statistics of fluid kinetic energy, dissipation rate, particle kinetic energy, and particle acceleration were evaluated. Different grid-based interpolation kernels, e.g., trilinear, fourth-order Lagrange, and cubic splines, for E2L and L2E interpolation were used. All these interpolation kernels have grid-based interpolation stencils, thus as the grid is refined and the particle-to-grid size ratio ($D_p/\Delta$) increases, the interpolation stencil becomes narrower, localizing the effect of the particle on the fluid and vice versa. It was shown that, in the absence of any correction model for self-disturbance, the fluid and particle statistics do not converge as the grid was refined. The fluid-particle interactions were better predicted when the grid is coarser (smaller $D_p/\Delta$), and hence neglecting the self-disturbance effect resulted in a smaller error on coarser grids. As the grid is refined, the self-disturbance field becomes stronger, because the particle reaction force is distributed over a smaller region due to localized interpolation kernels, resulting in a larger error.
In contrast, when a correction model was used to obtain an estimate of the undisturbed fluid velocity, consistent results were obtained for all kernels as the grid was refined, emphasizing the importance of self-disturbance correction, especially when the particle size is comparable to the grid size and grid-based interpolation kernels are used.



Particle size-based interpolation kernels, wherein the kernel width is proportional to the particle size, have also been used with Gaussian function~\citep{lomholt2002,deen2009direct,gualtieri2015,finn2016,vreman2016particle,fukada2018numerical,pakseresht2019}. These are commonly employed in dense
particle-laden flows, wherein the volumetric displacement by the presence of the particles is typically accounted for by using volume-filtered Navier-Stokes equations~\citep{apte2008,capecelatro2013,finn2016} to obtain a void fraction field smaller than unity.~\citet{vreman2016particle} investigated particle-turbulence interactions of 64 fixed particles in a statistically stationary homogenous isotropic turbulence using fully resolved direct numerical simulations, and used the data to compare predictions from point-particle models without any correction for the self-disturbance. The particle size was twice as large as the grid size. A simple top-hat interpolation kernel was used to distribute the force to the grid cells and varied the kernel width over $\frac{1}{2}D_p$, $2D_p$, and $4D_p$. The kernel width proportional to $4D_p$ was shown to be able to capture the point-particle model predictions on turbulence attenuation well in comparison to the fully resolved data. On the other hand, when the kernel width was smaller, the point-particle model underpredicted the turbulence attenuation. Without correcting for the self-disturbance, a kernel-width much larger than the particle size was able to sample the fluid velocity from the undisturbed region, resulting in better predictions. This study shows the effect of the particle size--based kernel widths in point-particle models, however, their impact in obtaining grid-convergent results has not been fully investigated. The top-hat kernel results in sudden changes in interpolation weights and is not appropriate for moving particles. The main advantage of the particle size--based kernels is that irrespective of the grid, the region of influence of the particle on the fluid remains the same. However, if the particle is comparable to the local grid size a kernel width larger than the particle size may require several neighbors of the control volume containing the particle, and hence such an interpolation kernel is computationally expensive, especially for complex, unstructured grids.


The main goal of the present work is to evaluate the accuracy and predictive capability of grid-based and particle size--based interpolation kernels for point-particle models with varying grid refinement, with and without accounting for the self-disturbance created by the particle. The present work focuses on two main hypotheses
when particle sizes are comparable to the grid size ($D_p\sim\Delta$). 
\begin{hyp}
The kernel width used in E2L and L2E interpolations should scale with the particle size and should be independent of the grid resolution. 
This keeps the region of influence of the particle the same, irrespective of the grid size.
\end{hyp}
\begin{hyp}
Using a kernel width much larger than the particle size for E2L interpolation  will sample the fluid velocity from a region with reduced influence from the self-disturbance field, thus improving the estimate of the fluid velocity at the particle location even without any correction model. However, a localized kernel {for L2E} with width about the size of the particle may be able to capture the reaction of the particle on the fluid more accurately.
\end{hyp}
The first hypothesis stems from the basic idea that the region of influence of the particle on the fluid flow should remain the same irrespective of the grid resolution employed. The second hypothesis suggests to use different kernel widths for the E2L and L2E interpolations. For E2L, if the kernel width is much larger than the particle size, the interpolation kernel will sample fluid velocities from a region that is not disturbed by the particle. On the other hand, if the particle force is distributed over the same large region, the local effect of the particle on the fluid may not be well captured. Hence, a localized kernel for L2E with width about the size of the particle may provide a better representation of the effect of the particle on the fluid flow.

To test these hypotheses, a detailed evaluation is conducted of the grid-based and particle size--based interpolation kernels and their widths with varying grid refinement. Two canonical test cases that of flow over a stationary particle, and particle-turbulence interaction in decaying, homogeneous isotropic turbulence at low volume loadings are studied.          Different interpolation kernels are used to quantify their effects on the predictive capability of the point-particle model with varying grid refinement, with and without the correction for self-disturbance. A grid resolution--based, compact, three-point \roma-delta function is given as
\begin{equation}
\label{eq:delta}
{\mathcal G}^{\sigma}(\mathbf{x}_{cv}-\mathbf{x}_p) =  \left\{ \begin{array}{ll}
			\frac{1}{6} (5-3|r| - \sqrt{-3(1-|r|)^2+1}),& ~~~~0.5\leq |r| \leq 1.5, \;r=|\mathbf{x}_{cv}-\mathbf{x}_p|/{\Delta} \\
			\frac{1}{3} (1+\sqrt{-3r^2+1}),& ~~~~|r| \leq 0.5 \\
			0, & ~~~~{\rm otherwise.}
			\end{array}
			\right.	
\end{equation}
This kernel is second-order, smoother than trilinear interpolation, and commonly used in immersed-boundary methods~\citep{roma1999adaptive}.
The second kernel is based on the particle size and is defined as a clipped Gaussian function,
\begin{align}
\label{eq:gaussian1}
&\mathcal{G}^{\sigma}(\mathbf{x}_{cv}-\mathbf{x}_p) =  \left\{ \begin{array}{ll}
			\frac{1}{\sigma \sqrt{2\pi}} \text{exp}\Big[-\Big(\frac{\mathbf{x}_{cv}-\mathbf{x}_p}{\sqrt{2}\sigma}\Big)^2\Big],& ~~~~|r| \leq 3\sigma, \;r=|\mathbf{x}_{cv}-\mathbf{x}_p| \\
			0, & ~~~~{\rm otherwise,}
			\end{array}
			\right. 
\end{align}
which provides smoother interpolations. The kernel width depends on a chosen standard deviation, $\sigma$, which scales with the particle diameter, i.e. $\sigma = cD_p$ where $c = \mathcal{O}(1)$. It is important to note that Gaussian functions are non-compact, and have long tails. In practice, however, the function is clipped once the weights become small, as in Eq.~\ref{eq:gaussian1}, and then the weights are normalized to enforce the conservation condition, $\int_{\mathcal V}{\mathcal G}^{\sigma}dV=1$. In a Gaussian kernel, 99.7\% of the interpolation weights are within $\pm 3\sigma$, thus choosing this as a cutoff point for the tails retains the majority of the kernel and ensures that the weights are very small at the filter cutoff, resulting in a smoother transition. Cutting off the filter tails too short would result in bigger discontinuities in interpolation weights at the cutoff point, possibly affecting the smoothness of particle and flow statistics. Two different kernel widths are used for this particle size--based kernel, corresponding to \gaussone ($\sigma = cD_p = \sqrt{2/\pi}D_p$) and \gausstwo ($\sigma = cD_p = 1.5D_p$). The \gaussone kernel is commonly used in force-coupling methods~\citep{lomholt2002}, wherein $\sigma$ is chosen such that the fluid velocity at the particle location matches the rigid-body motion of the particle, approximately enforcing a boundary condition at the particle location~\citep{gualtieri2015, maxey2001}. The \gausstwo kernel gives a width similar to the clipped fourth-order polynomial function used by~\citet{deen2009direct}.

\begin{figure}[!htpb!]
\begin{center}
\vspace{-3mm}
	\includegraphics[width=1\textwidth]{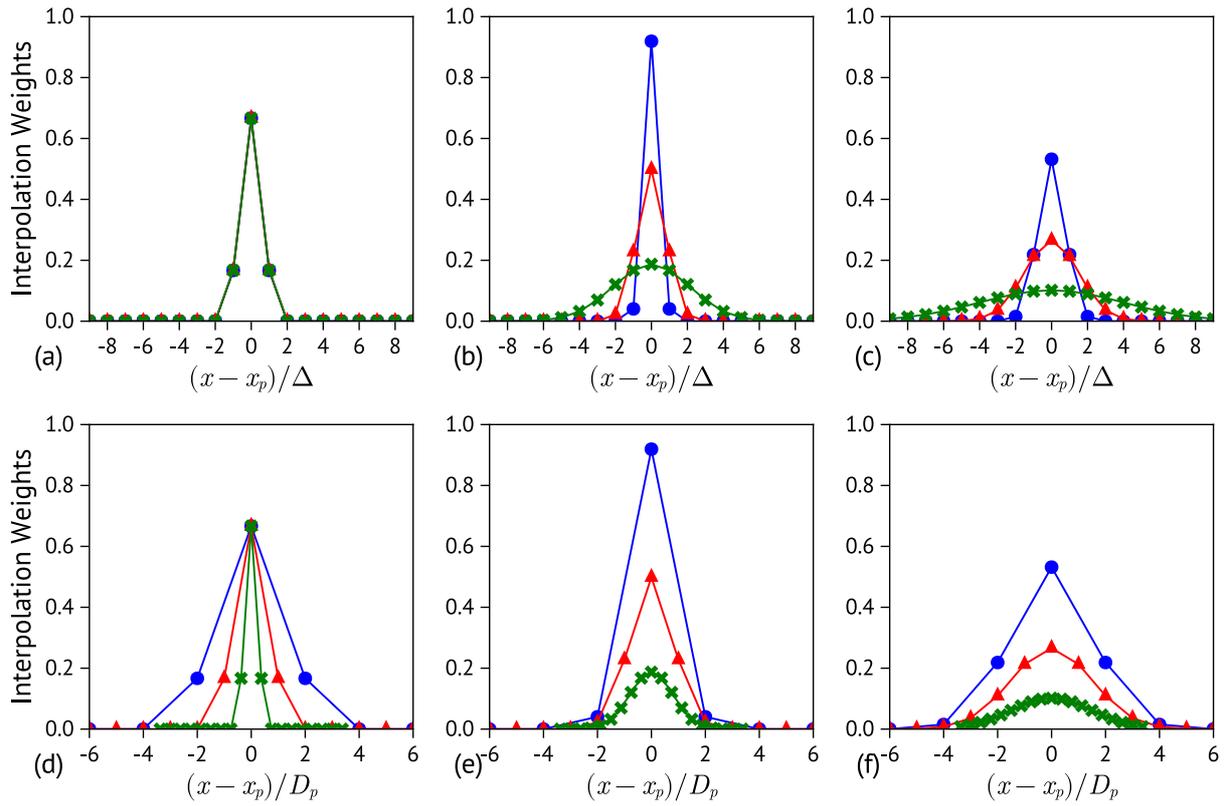}
 \vspace{-8mm}
     \caption{Interpolation weight distribution with grid refinement keeping the particle size unchanged: (a,d) \roma, (b,e) \gaussone, and (c,f) \gausstwo. The x-axis is normalized with respect to grid size (top panel, a--c) and  particle size (bottom panel, (d--f)). Different colors/symbols represent different grid resolutions: (\protect\bluecircle)~$D_p/\Delta=0.5$ (coarse), (\protect\redtriangle)~$D_p/\Delta=1.0$ (medium), and (\protect\greenx)~$D_p/\Delta=2.66$ (fine). Each marker represents the center of a control volume.}
     \label{fig:weights}
	\end{center}
\end{figure}

Figure~\ref{fig:weights} shows the comparison of weights under grid refinement for these three interpolation kernels, where each line marker represents the center of a control volume. In figures~\ref{fig:weights}(a-c), the $x$-axis is normalized by the grid size, $\Delta$, whereas in figures~\ref{fig:weights}(d-f), the $x$-axis is normalized by the particle size, $D_p$. It is useful to understand the extent of the interpolation kernel and its smoothness under grid refinement for a particle of a fixed size. As seen from figure~\ref{fig:weights}(a), under grid refined ($\Delta$ is decreased), the \roma-delta function has the same weights for the nearest neighbors of the control volume containing the particle. Since the grid is refined, this kernel distributes the weights in a narrower region, independent of the particle size to grid ratio. The particle size--based \gaussone and \gausstwo kernels, on the other hand, change the shape and weights under grid refinement as seen in figures~\ref{fig:weights}(b,c). However, the extent of the distribution of the weights is unchanged with varying grid refinement, as the particle size is fixed. Thus, for the finest grid, the kernel becomes smoother and distributes weights to several control volumes on either side of the particle. The weights under grid refinement and normalized by the fixed particle size are also informative. As the grid is refined, the \roma-delta kernel becomes narrower as shown in figure~\ref{fig:weights}(d), thus changing the region of influence of the particle under grid refinement. Whereas, the particle size--based \gaussone and \gausstwo kernels become smoother, and distribute the weights over the same region dictated by the kernel width, and keeping the region of influence of the particle the same.

It is thus clear that, for a grid-based kernel, the two-way coupling force distribution (L2E) and interpolation of the fluid velocity at the particle location (E2L) become highly localized as the grid is refined. This is true for any grid-based kernel, although the \roma-delta kernel is shown and used in this work. Such a kernel is sufficient when the particle is much smaller than the grid size ($D_p/\Delta << 1$); however, it can give rise to a large disturbance field for $D_p\sim \Delta$, as the reaction force from the particle is distributed in a very narrow region. On the other hand, for particle-size based \gaussone and \gausstwo kernels, the local weights become smaller under grid refinement, whereas the kernel distributes the reaction force over a region that scales with the particle size. Thus, for $D_p\sim \Delta$, the particle size--based kernel will distribute weights to several neighbors on each side of the particle, which can be computationally expensive for parallel implementations as well as for complex, unstructured grids, wherein a linked list of several neighboring cells needs to be carried. However, when the particle becomes much smaller than the grid size ($D_p/\Delta << 1$, not shown), the particle-size-based kernel will produce a sharp delta function at the particle location, resulting in zero or very small weights at the neighboring grid control volumes. This, however, is consistent with the fundamental assumptions employed in a point-particle approach. The effect of these kernels on the flow-particle interactions is clearly illustrated on a simple test case of uniform flow over a stationary particle in Section~\ref{subsec:flow_over_sphere}.

In the present work, evaluation of the point-particle models with and without self-disturbance correction is conducted under grid refinement, wherein the particle size becomes larger than the grid size. The self-disturbance correction is obtained by using an advection-diffusion-reaction (ADR) model equation for the disturbance field created by a particle and solved using the Zonal-ADR approach developed by~\citet{apteADR}. Results obtained from the grid-based \roma interpolation kernel are compared with those obtained from the particle size--based Gaussian kernels. The rest of the paper is arranged as follows. A brief description of the mathematical formulation for the disturbance field and its zonal implementation for each particle is given in Section~\ref{sec:methods}. Results for flow over a stationary particle are given in Section~\ref{subsec:flow_over_sphere}, followed by particle-laden decaying isotropic turbulence corresponding to the particle-resolved direct numerical simulation (PR-DNS) study of~\citet{mehrabadi2018direct}, at low Reynolds numbers and few particles, summarized in Section~\ref{subsubsec:low_hit}. The examination of particle-laden decaying isotropic turbulence is extended to a higher Reynolds number and a larger number of particles in Section~\ref{subsubsec:high_hit}. Lastly, conclusions are given in Section~\ref{sec:conclusions}. 


\section{Mathematical Formulation}
\label{sec:methods}
The mathematical formulation for the self-disturbance correction in the point-particle model for low-volume loadings is discussed in brief. The formulation is based on the incompressible Navier-Stokes equations,
\begin{eqnarray}
\label{eq:continuity}
\frac{\del  {u_j}}{\del x_j} &=&  0,\\
\label{eq:navier_stokes}
{\rho_g}\left(\frac{\del {u_i}}{\del t} + \frac{\del {u_j}{u_i} }{\del x_j} \right)&=& 
-\frac{\del{p}}{\del x_i} + {\mu}\frac{\del^2 u_i}{\del x_j^2} +  \underbrace{\left(-\sum_{q=1}^{N_p} {\mathcal G}^{\sigma}(\mathbf{x}_{cv}-\mathbf{x}_{q})F^t_{i,q}\right)}_{{\dot S_i}},
\end{eqnarray}
where ${\rho_g}$ is the density of the fluid, $\mu$ is the dynamic viscosity, $p$ is the pressure, and
${u}_i$ is the {two-way coupled} fluid velocity that includes the disturbances created by all $N_p$ particles through the net interphase source term ($\dot S_i$) based on all forces ($F_{i,q}^t$), except the gravity, acting on the particle located at ${\mathbf x}_q$ and projected onto the Eulerian grid control volumes located at ${\mathbf x}_{cv}$, using the interpolation kernel ${\mathcal G}^{\sigma}$, with $\sigma$ being the kernel width. 
The projection function satisfies the conservation condition, $\int_{\mathcal V}{\mathcal G}^{\sigma}d{\mathcal V} = 1$, where the integration is over the whole fluid volume ($\mathcal V$). 

For the point-particle model, the Lagrangian equations for the motion of the $p^{\rm th}$ particle are
\begin{eqnarray}
\label{eq:part}
\frac{dx_{i,p}}{dt} = u_{i,p},~~~ m_p\frac{du_{i,p}}{dt}=m_p\left(1 - \frac{\rho_g}{\rho_p}\right)g_i+ F_{i,p}^t,
\end{eqnarray}
where $x_{i,p}$ and $u_{i,p}$ are the particle position and velocity, respectively; $m_p$ is the mass of the particle, and $F_{i,p}^t$ represents forces acting on the particle including the drag, lift, added mass, pressure, and history, among others. In the present work, only the non-linear drag force is used.  The standard drag force on the particle $p$ can be modeled as,
\begin{equation}
F_{i,p}^{\rm drag} = m_p\frac{ u^{u,p}_{i,@x_p} - u_{i,p}}{\tau_r},~~ \tau_r = \frac{1}{f}\frac{\rho_p D_p^2}{18 \mu },~~~ f = 1 + 0.16 Re_p^{0.687},
\end{equation}
where $u^{u,p}_{i,@{\mathbf x}_p}$ represents the undisturbed fluid velocity seen by the particle $p$ and interpolated to the particle location (${\mathbf x}_p$), $D_p$ is the particle
diameter, $\tau_r$ is the particle relaxation time, and $f$ is the Schiller and Naumann nonlinear correction for drag coefficient based on the particle Reynolds number, \mbox{$Re_p = \rho_g D_p |u^{u,p}_{i,@{\mathbf x}_p} - u_{i,p}|/\mu$}.

The undisturbed flow field seen by a particle $p$ [denoted by superscript $(\cdot)^{u,p}$] can be obtained by excluding the reaction force from the $p^{\rm th}$ particle~\citep{apteADR},
\begin{eqnarray}
\label{eq:undisturbed1}
\frac{\del  {u_j}^{u,p}}{ \del x_j} &=&  0,\\
\label{eq:undisturbed2}
{\rho_g}\left(\frac{\del {u_i}^{u,p}}{\del t} + 
\frac{\del {u_i}^{u,p} {u_j}^{u,p}}{\del x_j}\right) &=& 
-\frac{\del{p}^{u,p}}{\del x_i} + {\mu}\frac{\del^2 {u_{i}^{u,p}}}{\del x_j^2}  - \sum_{q=1,q \ne p}^{N_p}{\mathcal G}^{\sigma}(\mathbf{x}_{cv}-\mathbf{x}_{q})F^t_{i,q}.
\end{eqnarray}
Subtracting Eqs.~(\ref{eq:continuity}-\ref{eq:navier_stokes}) from the corresponding Eqs.~(\ref{eq:undisturbed1}-\ref{eq:undisturbed2}), the self-disturbance field [denoted by superscript $(\cdot)^{d,p}$] created by the particle $p$ is given as
\begin{eqnarray}
\label{eq:cont_disturb}
\frac{\partial u_j^{d,p}}{\partial x_j}&=&0,\\
\label{eq:mom_disturb}
\rho_g\left(\frac{\partial u_i^{d,p}}{\partial t} +  \frac{\partial u_j  u_i^{d,p}}{\partial x_j}+  \frac{\partial  u_j^{d,p} u_i^{u,p}}{\partial x_j}\right)  &=& - \frac{\partial p^{d,p}}{\partial x_i} + {\mu}\frac{\del^2 {u_i^{d,p}}}{\del x_j^2} +
 {\mathcal G}^{\sigma}(\mathbf{x}_{cv}-\mathbf{x}_{q})F^t_{i,p},
\end{eqnarray}
where 
\begin{equation}
\label{eq:vel_und}
u_i^{d,p} = u_i^{u,p} - u_i,~~~
p^{d,p} = p^{u,p} - p.
\end{equation}
Note that in the self-disturbance equation, only the interaction force from the particle $p$ is needed. There is also nonlinear interaction between the disturbance field for particle $p$ and the two-way--coupled velocity ($u_i$) as well as the undisturbed velocity ($u_i^{u,p}$). 

~\citet{pakseresht2021disturbance} derived similar equations for a {\it single} particle system and showed accurate reconstruction of the undisturbed flow field. Solving the above set of equations for each particle in a multi-particle system, although possible, is expensive, as it requires the solution of a Poisson system for the pressure disturbance.~\citet{pakseresht2021disturbance} 
first suggested approximating the pressure and viscous terms in the disturbance equation as a diffusion term with effective viscosity $K_{\mu} \mu$. The idea for such approximation stems from the fact that in the Stokes flow limit, the pressure contribution to the drag force on a spherical particle is exactly half of the viscous contribution and has the same form as the viscous force. Hence, to match the drag force in the Stokes limit, $K_{\mu} = 1.5$ was used. In general, $K_{\mu}$ will vary based on the Reynolds number. Recent PR-DNS studies by~\citet{ganguli2019drag} showed that the ratio of pressure to viscous contribution to drag remains roughly the same up to a particle Reynolds number of 10. Hence, $K_{\mu}=1.5$ is used for all cases considered in this work, and good predictions up to  $Re_p = 100$, suggest that the assumption is reasonable even for large Reynolds numbers. The {approximate} disturbance equations for the particle $p$ then become
\begin{eqnarray}
\label{eq:navier_stokes2}
\rho_g\left(\frac{\partial u_i^{d,p}}{\partial t} + \frac{\partial u_j  u_i^{d,p}}{\partial x_j}+  \frac{\partial  u_j^{d,p} u_i^{u,p}}{\partial x_j}\right)  &=&  {K_{\mu}\mu}\frac{\del^2 {u_{i}^{d,p}}}{\del x_j^2} + 
 {\mathcal G}^{\sigma}(\mathbf{x}_{cv}-\mathbf{x}_{q})F^t_{i,p}.
\end{eqnarray}
The above nonlinear, unsteady advection-diffusion-reaction (ADR) equations are solved in addition to Eqs.~(\ref{eq:continuity}-\ref{eq:navier_stokes}), to obtain the undisturbed velocity from Eq.~(\ref{eq:vel_und}).

\subsection{The Zonal-ADR method}
\label{subsec:zonal}
The ADR equations for each particle $p$ only need the net force acting on the particle $F_{i,p}^t$ and can be solved separately. In addition, the undisturbed fluid velocity is  only needed at the particle location to compute the particle forces. These two aspects are exploited by solving the ADR equations in a small zone surrounding the particle in the Zonal-ADR approach. Details of the approach, numerical implementation, verification, and validation studies can be found in~\citet{apteADR} and only a brief summary is given here.

A Cartesian, collocated grid--based, second-order, fractional time--stepping solver has been developed \citep{finn2016} and used in the present work. An overset-grid algorithm is devised for the zonal solution and details of this numerical implementation can be found in~\citet{apteADR}. In the present case, the overset grid and the flow-solver grids are exactly aligned, and only the control volumes surrounding the particle location where the ADR equations need to be solved are tagged for each particle. As the particle moves, this region where the disturbance equation is solved is updated.  A zone containing $\pm cni$, $\pm cnj$, $\pm cnk$ control volumes is used, where $cni$, $cnj$, and $cnk$ correspond to the extent of the number of overset-grid points around the particle in $x$, $y$, and $z$ directions, respectively. Typically, the values of $cni$, $cnj$, and $cnk$ depend on the particle size as well as the local grid resolution. A typical value of $\pm6$--$8$ grid points is found to be sufficient for all the test cases studied in the present work.


The viscous and the other advection terms are approximated using the second-order Crank-Nicholson scheme and using the same spatial discretization algorithm as the baseline flow solver. The third term in the ADR Eq.~(\ref{eq:navier_stokes2}) represents the advection of the disturbance by the undisturbed flow velocity ($u_i^{u,p}$). This nonlinear term is treated explicitly by using the undisturbed flow velocity from the previous time-step. 
The disturbance field is generated by the particle reaction force and in the absence of any advection, it is simply diffused away from the particle. In the presence of advection, a convective outflow boundary condition is applied to boundaries of the overset grid that has outflux due to the two way--coupled velocity field, whereas no new disturbance is introduced at the influx boundaries. If there are walls present in the domain, the disturbance velocity field at the wall also goes to zero due to the no-slip condition. The Zonal-ADR approach requires the storage of the two-way coupled velocity field on the overset grid for each particle. Using only a few grid cells around the particle, the solution of the ADR equations is reasonably fast and its overhead is insignificant (around 20\%), with proper particle-load balancing.

\section{Results}
\label{sec:results}
To test Hypothesis 1 given in Section~\ref{sec: intro}, and to understand the effect of different kernels and kernel widths for E2L and L2E interpolations, a simple test case of a stationary particle in a uniform flow is investigated at parameters (Reynolds number, ratio of $D_p$ to $\Delta$) representative of those in the subsequent isotropic turbulence cases. 
The stationary particle test case is investigated with and without the Zonal-ADR correction, to quantify the effect of the kernels and kernel widths under grid refinement. 
Next, particle-turbulence interactions are investigated at low volume loadings in decaying isotropic turbulence laden with monodispersed, Kolmogorov-scale, spherical particles at low initial Reynolds number, $Re_{\lambda,0}=27$. The parameters are chosen corresponding to the particle-resolved DNS (PR-DNS) study by~\citet{mehrabadi2018direct} and the effect of different kernels and kernel widths under grid refinement is evaluated.


For most of these cases, the E2L and L2E interpolation functions are identical. To test Hypothesis 2, different E2L and L2E interpolation kernels with a narrower kernel of \roma-delta function for L2E and a wider kernel of \gausstwo for E2L are used (denoted as \romagausstwo). The effect of these non-symmetric interpolation kernels is again evaluated for the stationary particle case, as well as the decaying isotropic case.

Finally, taking the results from the lower Reynolds number case as validation and motivation for choice in interpolation kernels, the decaying isotropic turbulence cases are extended to a higher Reynolds number case with Kolmogorov-scale particles, with an order of magnitude more particles, two different Stokes numbers, and a range of grid sizes.

\subsection{Flow over a stationary sphere}
\label{subsec:flow_over_sphere}
In this section, flow over a stationary particle is investigated at a particle Reynolds number ($Re_p$) of 1, which is representative of the Reynolds numbers obtained in the isotropic turbulence case discussed later {in Section \ref{subsubsec:low_hit}}. A particle of size $D_p = 2\pi/96$ is placed at the center of a cubic domain of length $2\pi$. A grid size of $\Delta = D_p = 2\pi/96$ is used for baseline computations. A systematic grid refinement study, keeping the particle parameters the same, is carried out with grid sizes of $2\pi/48$, $2\pi/96$, $2\pi/192$, and $2\pi/256$. The main goal of this study is to quantify the effect of grid refinement on predicting the drag force on the particle using different interpolation kernels, namely, (i) \roma, (ii) \gaussone, (iii) \gausstwo, and (iv) \romagausstwo. The E2L and L2E interpolation functions are the same for the first three cases, whereas the \romagausstwo uses the \roma function for L2E and \gausstwo for E2L interpolations. 

\begin{table*}
\begin{center}
\begin{tabular}{|cc|ccc|ccc|}
\hline
\multirow{3}{*}{$D_p/\Delta$} & \multirow{3}{*}{Interpolation} & \multicolumn{3}{c|}{No Correction} & \multicolumn{3}{c|}{Zonal-ADR} \\ \cline{3-8} 
 &  & \multicolumn{2}{c}{\% Relative Error} & \multirow{2}{*}{$u_{@p}^{2w}/u_{in}$} & \multicolumn{2}{c}{\% Relative Error} & \multirow{2}{*}{$u_{@p}^{2w}/u_{in}$} \\ \cline{3-4} \cline{6-7}
 &  & $u_{@p}^{un}$ & $\|F_p\|$ &  & $u_{@p}^{un}$ & $\|F_p\|$ &  \\ 
 \hline
 \hline
0.5 & \roma & 16.3 & 17.56 & 0.84 & 1.24 & 1.36 & 0.798 \\
0.5 & \gaussone & 16.3 & 17.5 & 0.837 & 0.83 & 1.35 & 0.882 \\
0.5 & \gausstwo & 10.4 & 11.2 & 0.897 & 0.35 & 0.9 & 0.87 \\
0.5 & \romagausstwo & 13.0 & 13.78 & 0.872 & 0.59 & 0.63 & 0.81 \\ \hline
1 & \roma & 33.7 & 35.8 & 0.66 & 0.87 & 0.9 & 0.47 \\
1 & \gaussone & 22.5 & 24.6 & 0.76 & 0.6 & 0.67 & 0.7 \\
1 & \gausstwo & 11.5 & 12.5 & 0.88 & 0.35 & 0.38 & 0.87 \\
1 & \romagausstwo & 16.5 & 17.7 & 0.84 & 0.59 & 0.63 & 0.81 \\ \hline
2 & \roma & 53.4 & 55.9 & 0.47 & 5.4 & 5.8 & -0.16 \\
2 & \gaussone & 23.1 & 24.6 & 0.77 & 0.1 & 0.1 & 0.3 \\
2 & \gausstwo & 13.8 & 14.9 & 0.86 & 0.9 & 0.98 & 0.83 \\
2 & \romagausstwo & 19.6 & 21.3 & 0.8 & 0.13 & 0.16 & 0.25 \\ \hline
2.66 & \roma & 60.9 & 63.32 & 0.39 & 7.1 & 7.6 & -0.558 \\
2.66 & \gaussone & 16.3 & 17.5 & 0.837 & 1.47 & 1.6 & 0.625 \\
2.66 & \gausstwo & 10.3 & 11.2 & 0.897 & 0.97 & 1.05 & 0.82 \\
2.66 & \romagausstwo & 12.8 & 13.7 & 0.872 & 1.4 & 1.3 & 0.724 \\ \hline
\end{tabular}
\caption{Effect of grid refinement on undisturbed fluid velocity, magnitude of particle drag force, and the disturbed fluid velocity at the particle location for different interpolation kernels.\label{tab:stationary}}
\end{center}
\end{table*}


Table~\ref{tab:stationary} shows the relative error in computing the undisturbed fluid velocity at the particle location ($u^{un}_{@p}$) with and without the Zonal-ADR correction, compared to the true value [corresponding to the inlet velocity ($u_{\text{in}}$) of 1], the relative error in the particle drag force $F_p$, and the actual disturbed two way--coupled velocity at the particle location normalized by the inlet velocity ($u^{2w}_{@p}/u_{\text{in}}$). Without the Zonal-ADR correction, the errors in particle force and the undisturbed fluid velocity at the particle location are large (10--60\%) and {the errors} get worse with grid refinement, especially when the grid-based \roma-delta function is used. This result is consistent with the main conclusions of~\citet{horwitz2020}, who used trilinear interpolation for particle-laden isotropic turbulence and observed that coarser meshes resulted in better predictions and the prediction errors became worse with grid refinement. As the grid is refined, the \roma kernel becomes narrower and distributes the particle reaction force to the nearest neighbors of the control volume containing the particle (see also figures~\ref{fig:weights}[a,d]). This creates a strong disturbance field and, without any correction, results in a large error. With the Zonal-ADR correction, even for a grid-based \roma interpolation kernel, particle force, and undisturbed fluid velocity errors are significantly smaller for all grid refinements. However, when the grid resolution is much finer than the particle size ($D_p/\Delta=2.66$), a negative two-way fluid velocity at the particle location is observed, which is unphysical based on the flow around a particle obtained from particle-resolved simulations. This suggests that, with a very narrow interpolation kernel, a large particle reaction force creates an extremely strong disturbance resulting in negative fluid velocity at the particle location. With correction, the reaction force, $F_p$, is larger compared to without correction, as the correction scheme provides the undisturbed fluid velocity at the particle location reliably. Even though the Zonal-ADR correction scheme reduces the error in the undisturbed fluid velocity, the two-way coupled velocity field at the particle location goes in the reverse direction, for particles much larger than the grid. Without correction, however, for these large particles, the two-way coupled velocity at the particle location is in the main flow direction, because the reaction force is underpredicted. Overall, this suggests that the two-way coupled velocity will be incorrect or unphysical both with and without correction when using a \roma kernel and when the particle size is larger than the grid size. This suggests that
 the grid-based kernel should not be used when the particle size is larger than the grid size.


On the other hand, when the interpolation kernel width is based on the particle size (\gaussone and \gausstwo), the errors are reduced even when no correction is used and they remain small with varying grid refinement. With the Zonal-ADR correction, the errors are significantly lower ($<$ 1.5\%) compared to the \roma interpolation, and the two-way fluid velocity at the particle location remains physical and does not become negative. Interestingly, even the \romagausstwo (L2E-E2L) interpolation results in small errors compared to the \roma (L2E-E2L) kernel, even without the Zonal-ADR correction, and the errors are comparable to the \gaussone or \gausstwo kernels for L2E-E2L interpolation without correction. This suggests that, without correction, using a larger kernel width based on particle size for E2L and a narrower kernel for L2E distribution provides a reasonable approximation, as hypothesized in Hypothesis 2. 
 This suggests that without correction, using a wider stencil, proportional to the particle size, to obtain the fluid velocity at the particle location samples the flow from a region less affected by the self-disturbance of the particle, resulting in better predictions. Numerical errors produced by not using the same interpolation stencils for L2E and E2L~\citep{sundaram1996numerical} are compensated by better prediction of the undisturbed fluid velocity using particle-based kernels, as postulated in Hypothesis 2. However, symmetric (E2L-L2E) kernels, such as \gaussone and \gausstwo, generally give smaller errors. With correction, the non-symmetric or symmetric particle-based kernels provide similar results, and the errors are still much lower than without correction.
 

\subsection{Decaying isotropic turbulence}
\label{subsec:decaying_HIT}
In this section, decaying isotropic turbulence laden with Kolmogorov-scale particles is investigated with and without the self-disturbance correction under systematic grid refinement for the different interpolation kernels. First, a low Reynolds number study is performed corresponding to the particle-resolved DNS of~\citet{mehrabadi2018direct} for validation, followed by a higher Reynolds number case with a large number of particles.

\subsubsection{Lower Reynolds number validation case}
\label{subsubsec:low_hit}

\begin{table*}
\begin{center}
\begin{tabular}{|c|c|c|}
\hline
 & Validation case & Higher Reynolds number case \\
 \hline
$N^3$ & $96^3$, $144^3$, $192^3$, $256^3$ & $128^3$, $192^3$, $256^3$, $384^3$, $512^3$ \\
$L$ & $2\pi$ & $2\pi$ \\
$\mu$ & 0.02058 & 0.005245 \\
$k_{max}\eta_0$ & 3.14, 4.7, 6.28, 8.36 & 1.5, 2.25, 3.0, 4.5, 6 \\
$k_{f,0}$ &1.0045, 1.0335, 1.0335, 1.0335 & 1.0008 \\
$\epsilon_{f,0}$ &0.4169, 0.46086, 0.46659, 0.42697 & 0.4463, 0.4564, 0.4618, 0.4651, 0.4665 \\
$c_{L}$ & 0.415 & 1.212, 1.213, 1.832, 1.686, 1.772 \\
$c_{\eta}$ & 0.4 & 0.419, 0.417, 0.420, 0.420, 0.420 \\
$Re_{\lambda,0}$ & 27 & 53.4, 52.8, 52.5, 52.3, 52.2 \\
$D_p/\Delta$ & 1, 1.5, 2, 2.66 & 0.477, 0.716, 0.955, 1.432, 1.91 \\
$St_{\eta, 0}$ & 100 & 10, 100 \\
$\rho_p/\rho_f$ & 1800 & 180, 1800 \\
$N_p$ & 1689 & 36796 \\
$\phi$ & 0.001 & 0.001 \\
$D_p/\eta_0$ & 1.0 & 1.0 \\
$\phi_m$ & 1.8 & 0.18, 1.8 \\
\hline
\end{tabular}
\caption{Initial flow and particle parameters corresponding to the different decaying isotropic turbulence cases.}
\label{tab:decayingHIT_parameters}
\end{center}
\end{table*}

Particle-laden, decaying, isotropic turbulence corresponding to the particle-resolved data by~\citet{mehrabadi2018direct}, at Taylor microscale Reynolds number of $Re_{\lambda}\approx 27$, is investigated. The computational domain is a triply periodic cubic box of side length $2\pi$. The initial condition for each case is the divergence-free random field sampled from Pope's model energy spectrum~\citep{pope2000turbulent}. The initial Kolmogorov length scale, $\eta_0$, is set to $2\pi/96$. The initial condition for each simulation is a divergence-free random field whose energy spectrum obeys Pope’s model spectrum,
\begin{equation}
E(\kappa) = C \epsilon^{2/3} \kappa^{-5/3} f_L(\kappa L)f_{\eta}(\kappa \eta),
\end{equation}
where $C=1.5$ is a model constant, $\kappa$ is the wavenumber, $\epsilon$ is the dissipation rate, and $L$ and $\eta$ are large eddy and Kolmogorov length scales, respectively. The functions  $f_L$  and  $f_{\eta}$  determine the shape of energy-containing and dissipative range of the energy spectrum and are defined as,
\begin{eqnarray}
f_L(\kappa L) &=& \left[\frac{\kappa L}{[(\kappa L)^2+c_L]^{1/2}} \right]^{5/3+p_0} \\
f_{\eta}(\kappa \eta)&=& {\rm exp}\left[- \beta\left([(\kappa \eta)^4 +c_{\eta}^4]^{1/4}-c_{\eta}\right)\right].
\end{eqnarray}
The model constants $p_0=2.0$ and $\beta=5.2$ are the same as suggested by~\citep{pope2000turbulent}, while $c_L$ and $c_{\eta}$ are determined to match the energy and dissipation rate required for the chosen $Re_{\lambda,0}$ and $D_p = \eta_0$. These constants change for different grid resolutions and grid arrangements (collocated used in the present study). Apart from $c_L$, $c_{\eta}$, $k_{f,0}$, and $\eta_{f,0}$, all other parameters for the model spectrum are similar to those used by~\citet{mehrabadi2018direct}. 

A summary of the simulation parameters in this section is given in Table \ref{tab:decayingHIT_parameters}. For the baseline case, the grid size ($\Delta$) is selected to be the same as the initial Kolmogorov scale, which is equal to the particle size ($D_p=\eta_0$). The particle-to-fluid density ratio is $\rho_p/\rho_f = 1800$, and the volume and mass loading are $\phi = 0.001$ and $\phi_m = 1.8$, respectively, giving a total  of $N_p = 1689$ particles in the domain. This results in a large particle Stokes number, $St_{\eta,0} = (1/18)(\rho_p/\rho_f)(D_p/\eta_0)^2= 100$. Particle dynamics is based only on the drag force modeled using the standard Schiller-Naumann drag correlation. Keeping all the above parameters the same, the grid is systematically refined ($\Delta = 2\pi/96$, $2\pi/144$, $2\pi/192$, and $2\pi/256$), and simulations are carried out for the four different interpolation kernels described before. Thus, $D_p/\Delta$ ranges between $1$--$2.66$ from the coarsest to the finest resolution. Following~\citet{mehrabadi2018direct}, particles are injected at random positions with the initialization of the flow field using the Pope spectrum, and the particle velocity is set equal to the fluid velocity interpolated to the particle location using trilinear interpolation. For all cases, the magnitude of the velocity derivative skewness of these cases rapidly approaches a value of around 0.5 in about 0.1 turnover times.

\begin{figure}
\begin{center}
    \includegraphics[width=1\textwidth]{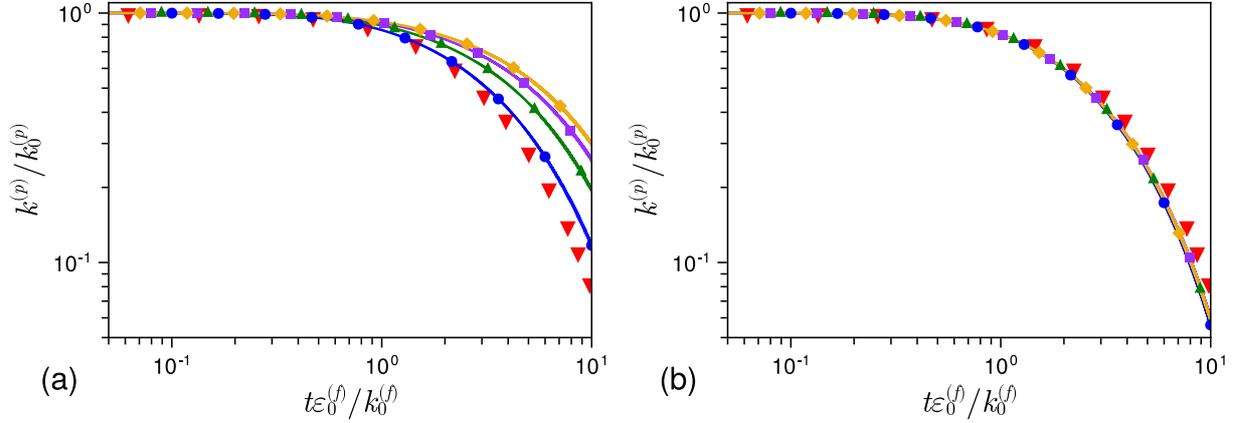}
    \caption{Temporal evolution of normalized particle kinetic energy with the grid-based \roma kernel for (a) no model, and (b) Zonal-ADR, under grid refinement: (\protect\bluecircle)~$D_p/\Delta=1$, (\protect\greentriangle)~$D_p/\Delta=1.5$, (\protect\purplesquare)~$D_p/\Delta=2$, and (\protect\golddiamond)~$D_p/\Delta=2.66$. The PR-DNS data from~\citet{mehrabadi2018direct}~(\protect\reddowntriangle) is also shown for comparison. }
    \label{fig:KEPEPSRoma}
\end{center}
\end{figure}

Figure~\ref{fig:KEPEPSRoma}(a,b) shows the temporal evolution of particle kinetic energy normalized by the initial fluid phase kinetic energy obtained using the grid-based \roma interpolation kernel for four different grid sizes with and without any self-disturbance correction. Without any correction model, as the grid is refined, the prediction of particle kinetic energy becomes more inaccurate, a result consistent both with the stationary particle test case  presented {in Section \ref{subsec:flow_over_sphere}} and with the conclusions of~\citet{horwitz2020}. As the grid is refined, the \roma interpolation kernel becomes narrow, adding a large reaction force in the neighborhood of the particle and creating a strong disturbance field. Since the fluid velocity is sampled using the same interpolation kernel and without any correction, the particle force is underpredicted, resulting in the slower decay of the particle kinetic energy. With the Zonal-ADR correction, however, all grid resolutions predict the temporal evolution of the particle kinetic energy very similar to the PR-DNS data. Thus, even for this grid-based, narrow interpolation kernel, the Zonal-ADR correction captures the fluid-particle interactions fairly accurately. The small mismatch compared to the PR-DNS data in Figure~\ref{fig:KEPEPSRoma}(b) may be attributed to the lack of knowledge of all the exact parameters for the initial spectrum used in PR-DNS, and the use of a collocated grid-based solver in the present work. Although the energetics of the particle-fluid interactions are captured fairly well with the correction scheme, the grid-based interpolation kernel can add large reaction force for particles larger than the grid size ($D_p/\Delta > 1$) and result in unphysical two-way coupled velocity field (as was seen in the stationary particle simulation in Section \ref{subsec:flow_over_sphere}). This could adversely affect particle and fluid statistics, e.g., see discussion on particle acceleration statistics in Section \ref{subsubsec:high_hit}.

\begin{figure}[!htpb!]
         \vspace{-4mm}
\begin{center}
    \includegraphics[width=1\textwidth]{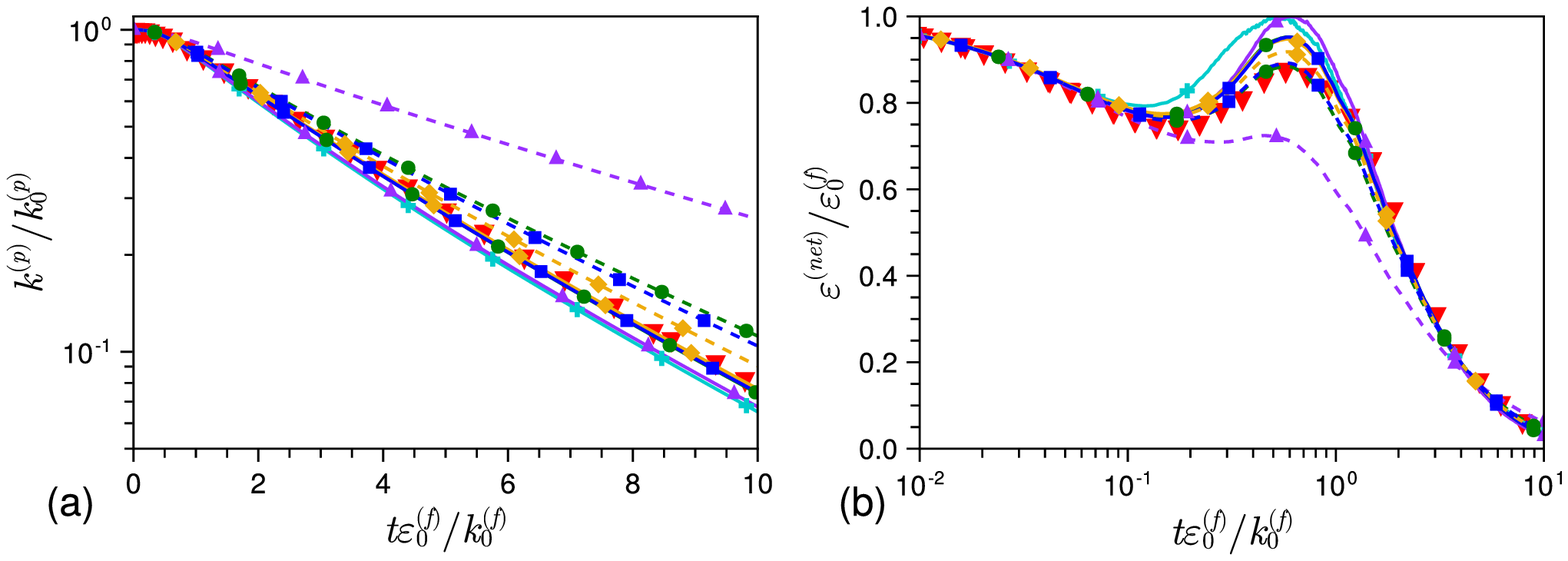}
         \vspace{-4mm}
     \caption{Temporal evolution of (a) normalized particle kinetic energy and (b) net dissipation rate using the Zonal-ADR model (solid lines) and no model (dashed lines) with different interpolation kernels on the $192^3$ grid: (\protect\purpletriangle)~\roma, (\protect\greencircle)~\gaussone, (\protect\golddiamond)~\gausstwo, and (\protect\bluesquare)~\romagausstwo. The PR-DNS data of~\citet{mehrabadi2018direct}~(\protect\reddowntriangle) and predictions using the \citet{esmaily2018} correction scheme with trilinear interpolation implemented in the present solver~(\protect\cyanx) is also shown for comparison. The \gaussone and \gausstwo lines for Zonal-ADR nearly overlap. }
     \label{fig:KEP_EPS_withEH}
	\end{center}
\end{figure}

The effect of using interpolation kernel widths proportional to the particle size is investigated next. Figure~\ref{fig:KEP_EPS_withEH}(a) shows the temporal evolution of normalized particle kinetic energy using the four interpolation kernels with and without the correction scheme on the $192^3$ grid. Even without any correction, the kinetic energy decay rate is reasonably well captured by the \gaussone, \gausstwo, and \romagausstwo interpolation kernels. The kinetic energy is slightly overpredicted without correction. This is because the fluid velocity at the particle location contains the self-disturbance and results in a smaller particle force, similar to the stationary particle case. With correction, however, the results follow the PR-DNS study reasonably well for all kernels, with \gausstwo being closest to the PR-DNS, whereas the grid-based \roma kernel gives a larger deviation. Table~\ref{tab:Kep} documents the particle kinetic energy at different times with and without the correction model for different interpolation kernels compared to the PR-DNS study. Also shown is the PP-DNS study conducted by~\citet{mehrabadi2018direct}, using the correction scheme developed by~\citet{horwitz2018}, and predictions from the E\&H correction scheme~\citep{esmaily2018} with trilinear interpolation implemented in the present solver. 

\begin{table*}
\begin{center}
\begin{tabular}{|c|ccc|ccc|ccc|}
\hline
\multirow{2}{*}{$\frac{t\epsilon_0^{(f)}}{k_0^{(f)}}$} & \multirow{2}{*}{PR-DNS} & \multirow{2}{*}{\begin{tabular}[c]{@{}c@{}}PP-DNS\\ (Trilinear)\end{tabular}} & \multirow{2}{*}{\begin{tabular}[c]{@{}c@{}}E\&H\\ (Trilinear)\end{tabular}} & \multicolumn{3}{c|}{No Correction} & \multicolumn{3}{c|}{Zonal-ADR} \\ \cline{5-10} 
 &  &  &  & \roma & \begin{tabular}[c]{@{}c@{}}\gaussone\\ $\frac{\sigma}{D_p}=0.8$\end{tabular} & \begin{tabular}[c]{@{}c@{}}\gausstwo\\ 1.5\end{tabular} & \roma & \begin{tabular}[c]{@{}c@{}}\gaussone\\ $\frac{\sigma}{D_p}=0.8$\end{tabular} & \begin{tabular}[c]{@{}c@{}}\gausstwo\\ 1.5\end{tabular} \\ \hline
0.54 & 0.9303 & 0.9267 & 0.9246 & 0.9645 & 0.9410 & 0.9351 & 0.9236 & 0.9283 & 0.9279 \\
2.7 & 0.5172 & 0.4732 & 0.4701 & 0.7028 & 0.5562 & 0.5272 & 0.4693 & 0.4977 & 0.5010 \\
4.87 & 0.2855 & 0.2423 & 0.2453 & 0.5141 & 0.3331 & 0.3004 & 0.2483 & 0.2670 & 0.2743 \\
6.55 & 0.1821 & - & 0.1562 & 0.4081 & 0.2302 & 0.1999 & 0.1564 & 0.1714 & 0.1754 \\ \hline
\end{tabular}
\caption{Normalized particle kinetic energy ($k^{(p)}/k_0^{(f)}$) at different times predicted with and without the Zonal-ADR model with different interpolation kernels and compared against PR-DNS as well as PP-DNS by~\citet{mehrabadi2018direct}. E\&H represents the implementation of the correction scheme by~\citet{esmaily2018} in the present solver.\label{tab:Kep}}
\end{center}
\end{table*}

Figure~\ref{fig:KEP_EPS_withEH}(b) shows the temporal evolution of the net dissipation rate normalized by the initial dissipation rate for the various interpolation kernels with and without correction. As shown by~\citet{sundaram1996numerical} and~\citet{mehrabadi2018direct}, the evolution equation for the mixture kinetic energy of the system $e_m = (1-\phi)\rho_fk^{(f)} + \phi \rho_p k^{(p)}$ is
\begin{equation}
\label{eq:dissipation}
\frac{de_m}{dt} = \underbrace{(1-\phi)\frac{1}{V}\int_V \mu {\mathbf u}_f \nabla^2{\mathbf u}_f dV}_{-\epsilon^{(f)}} + \underbrace{\frac{1}{V}\sum_{i=1}^{N_p} F_i \cdot (u_{p,i})}_{\Pi^{(p)}}-\underbrace{(1-\phi)\frac{1}{V}\sum_{i=1}^{N_p} F_i \cdot (u_{f@p,i})}_{\Pi^{(f)}},
\end{equation}
where $F_i$ is the force acting on a particle, $-\epsilon^{(f)}$ is the kinetic energy dissipation rate resolved on the grid, $\Pi^{(p)}$ is the particle kinetic energy dissipation rate, 
and $\Pi^{(f)}$ is the interphase kinetic energy transfer term between the particle and fluid 
Here, $-\epsilon^{(*)} = \Pi^{(p)} - \Pi^{(f)}$ represents the additional  dissipation near the particle surfaces~\citep{sundaram1996numerical}. In the present work, the net dissipation rate [$\epsilon^{(net)}=-\epsilon^{(f)} + \Pi^{(p)} - \Pi^{(f)}$] is computed from the rate of change of the fluid and particle kinetic energy [left-hand side of Eq.~(\ref{eq:dissipation})] for the numerical accuracy reasons. Without correction, the grid-based \roma interpolation significantly underpredicts the net dissipation rate. However, particle size--based interpolation kernels capture the trends of the PR-DNS data even without correction. The peak in the dissipation rate, created mainly by the no-slip conditions and resultant flow disturbance in PR-DNS, is also well captured by the point-particle model with correction. The slight overprediction in peak compared to the PR-DNS is attributed to not having the same exact initial conditions as in the PR-DNS. This is also confirmed by implementing a different correction scheme~\citep{esmaily2018}, which uses trilinear interpolation, into the same flow solver which shows a similar peak in dissipation rate with correction.
\begin{figure}
\begin{center}
	\includegraphics[width = 0.6\textwidth]{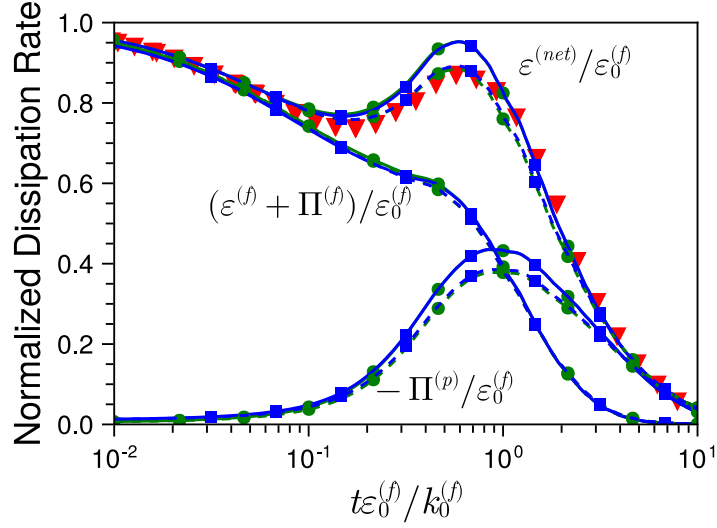}
    \caption{Temporal evolution of the resolved fluid (-$\epsilon^{(f)}$), particle dissipation rates ($\Pi^{(p)}$), and interphase energy transfer ($-\Pi^{(f)}$) to the 
    net normalized dissipation rate ($\epsilon^{(net)}/\epsilon_0^{(f)}$) using the Zonal-ADR correction (solid lines) and no model (dashed lines) with different interpolation kernels: (\protect\greencircle)~\gaussone and (\protect\bluesquare)~\romagausstwo. The PR-DNS data of~\citet{mehrabadi2018direct}~(\protect\reddowntriangle) is also shown for comparison.}
    \label{fig:EPS}
	\end{center}
\end{figure}
Figure~\ref{fig:EPS} shows the contributions of the resolved dissipation rate and interphase energy transfer [$-\epsilon^{(f)} - \Pi^{(f)}$], the particle kinetic energy dissipation rate to the net dissipation rate for two interpolation kernels (\gaussone and \romagausstwo) with and without correction. All parts of the dissipation rate are well captured by the particle-based interpolation kernel, even with the asymmetric \romagausstwo interpolation. This suggests that sampling fluid velocity from a region less perturbed by the self-disturbance of the particle compensates for the errors introduced by nonsymmetric interpolation in the kinetic energy conservation, especially when the particles are on the order of the grid size. 

\subsubsection{Higher Reynolds number case}
\label{subsubsec:high_hit}
The prior cases allowed testing of Hypotheses 1 and 2 against a simple, uniform flow over a stationary particle as well as decaying isotropic turbulence laden with Kolmogorov-scale particles for which PR-DNS data are available.
However, due to the low Reynolds number and maintaining of particle size equal to the Kolmogorov scale, those results were obtained with a small number of particles. In this section, Hypothesis 1 is tested for higher Reynolds number, $Re_{\lambda, 0} \approx 52$, two different Stokes numbers, and a larger number of particles and yet maintaining low volume loading. This Reynolds number is still somewhat limited by the desire to model Kolmogorov-scale particles and maintain a reasonable number of grid points under grid refinement. Test cases were designed to sufficiently resolve the small scales, ensuring $k_{max}\eta > 1$, where $k_{max} = \pi/\Delta$~\citep{Balachandar.Maxey1989, Yeung.Pope1988}. In particular, the most restrictive, i.e. coarsest case has $k_{max}\eta = 1.5$. The higher Reynolds number allows many more Kolmogorov-scale particles while maintaining low-volume loading. Under grid refinement, a wide range of particle size to grid size ($D_p/\Delta$) is studied to test Hypothesis 1.

A summary of the simulations in this section is given in Table \ref{tab:decayingHIT_parameters}. The domain is once again a triply periodic cube of side length $2\pi$, with initial flow conditions for each case coming from a divergence-free random field sampled from Pope's model energy spectrum~\citep{pope2000turbulent}. Five grid resolutions are chosen to span a range of particle-to-grid size ratios. The particle size is chosen to equal the initial Kolmogorov scale ($D_p = \eta_0$), which is held constant across all five grid refinements ($\Delta = 2\pi/128$, $2\pi/192$, $2\pi/256$, $2\pi/384$, and $2\pi/512$), resulting in a particle-to-grid size ratio ($D_p/\Delta$) which spans a range from $0.477$--$1.91$. Two different particle-to-fluid density ratios are tested ($\rho_p/\rho_f = 180$ and 1800) in order to achieve particle Stokes numbers of $St_{\eta,0} = 10$ and 100, respectively. In all cases, a volume fraction of $\phi = 0.001$ is used, resulting in mass loading of $\phi_m = 0.18$ and 1.8, respectively for the two Stokes numbers, giving a total of $N_p = 36796$ particles in the domain. Again, particle dynamics is based only on the drag force modeled using the standard Schiller-Naumann drag correlation. 

In the lower Reynolds number cases, validating against the PR-DNS of~\citet{mehrabadi2018direct}, the \gaussone and \romagausstwo cases provide similar results and both compare favorably to the PR-DNS results when combined with the Zonal-ADR correction. The wider \gausstwo kernel produced statistics slightly closer to the PR-DNS, especially in the uncorrected case due to its wide region of influence, but differences in the corrected results are small. Due to its wider stencil, the \gausstwo kernel requires more neighboring cells for larger particles, increasing the computational cost. Thus, opting for both consistency between E2L and L2E interpolation kernels (that results in energy conservation) and for a balance between accuracy and computational efficiency, the higher Reynolds number cases are only run for the grid-based \roma interpolation kernel and the particle size--based \gaussone interpolation kernel. Particles are initially injected at random positions and the initial particle velocity is set equal to the fluid velocity interpolated to the particle location using trilinear interpolation. Similar to the low Reynolds number cases, particles are injected at the start of decay. The magnitude of the velocity derivative skewness of these cases rapidly approaches a value of around 0.5 in about 0.1 turnover times.

The general results in the $St=10$ follow those from the $St=100$ case. In order to keep the same number of particles and volume loading between cases, the mass loading decreases to 0.18 for the $St=10$ case, resulting in the particle phase having a much weaker overall effect on the fluid phase. For this reason, some of the differences between cases are smaller, however, the main patterns remain consistent with the $St=100$ case, and hence some of the results for the $St=10$ case have been omitted for brevity.

\begin{figure}
    \centering
    \includegraphics[width=1.0\textwidth]{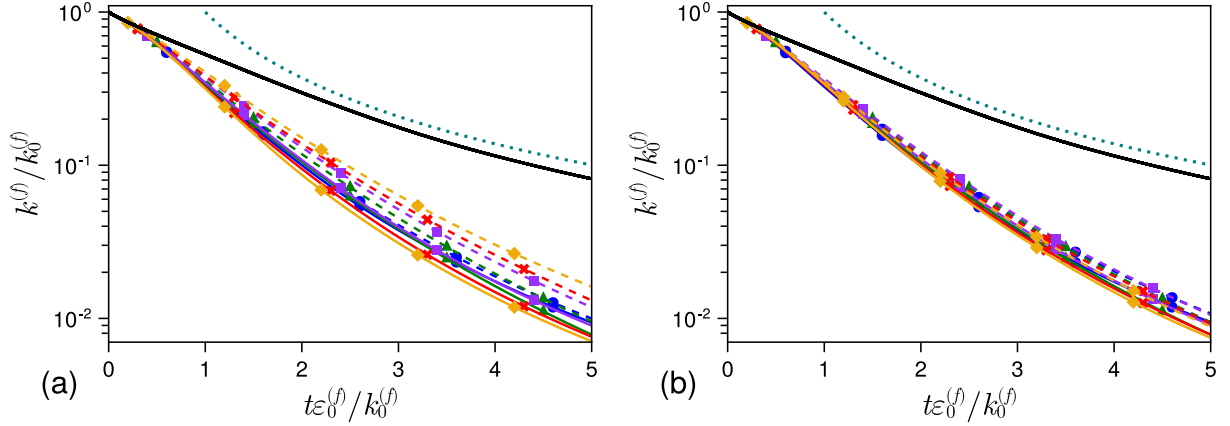}
    \caption{Temporal evolution of normalized fluid kinetic energy decay in the $Re_{\lambda,0} = 52$, $St=100$ case using the (a) grid-based \roma interpolation kernel and (b) particle size--based \gaussone interpolation kernel. Solid lines represent cases with the Zonal-ADR model while dashed lines represent the uncorrected cases. Different colors/symbols represent different grid sizes: (\protect\bluecircle)~$D_p/\Delta=0.477$, (\protect\greentriangle)~$D_p/\Delta=0.716$, (\protect\purplesquare)~$D_p/\Delta=0.955$, (\protect\redx)~$D_p/\Delta=1.432$, (\protect\golddiamond)~$D_p/\Delta=1.91$. Kolmogorov's selfsimilar decay rate,~$t^{-10/7}$~(\protect\tealdotted{}), is shown as a reference for the unladen case~(\protect\blackline{}).}
    \label{fig:Re52_KEf_St100}
\end{figure}

Figure \ref{fig:Re52_KEf_St100}(a,b) shows the temporal evolution of fluid kinetic energy for $St=100$ normalized by its initial value for the corrected and uncorrected cases calculated using the (i) grid-based \roma and (ii) particle size--based \gaussone interpolation kernels for all five grid refinements. The self-similar Kolmogorov decay rate, $t^{-10/7}$,~\citep{Kolmogorov1941} is shown as a reference against the corresponding decay rate of the unladen case. The fluid kinetic energy calculated with the uncorrected, grid-based \roma interpolation kernel diverges as the grid is refined, with the rate of decay decreasing as the grid size is decreased. By employing the Zonal-ADR correction scheme with \roma interpolation, the fluid kinetic energy is much more grid-converged, with very slight differences possibly due to slight differences in the initial fields for the different grid resolutions. When the same flow realizations are instead run with the particle size--based \gaussone interpolation kernel, even the uncorrected cases exhibit grid convergence across all resolutions. Again, there are slight variations at later times, but not as definitive as in the corresponding grid-based \roma interpolation. In general, even with the \gaussone interpolation scheme, the uncorrected cases result in slightly less fluid kinetic energy decay than their Zonal-ADR corrected counterparts, although since the volume loading in these simulations is fairly dilute, the differences in observed fluid kinetic energy are small. The effect of correcting for the self-disturbance field created by a particle is much more profound on particle statistics themselves and could be expected to have more effect on fluid statistics in a heavier volume loading than what is studied here.

\begin{figure}
    \centering
    \includegraphics[width=1.0\textwidth]{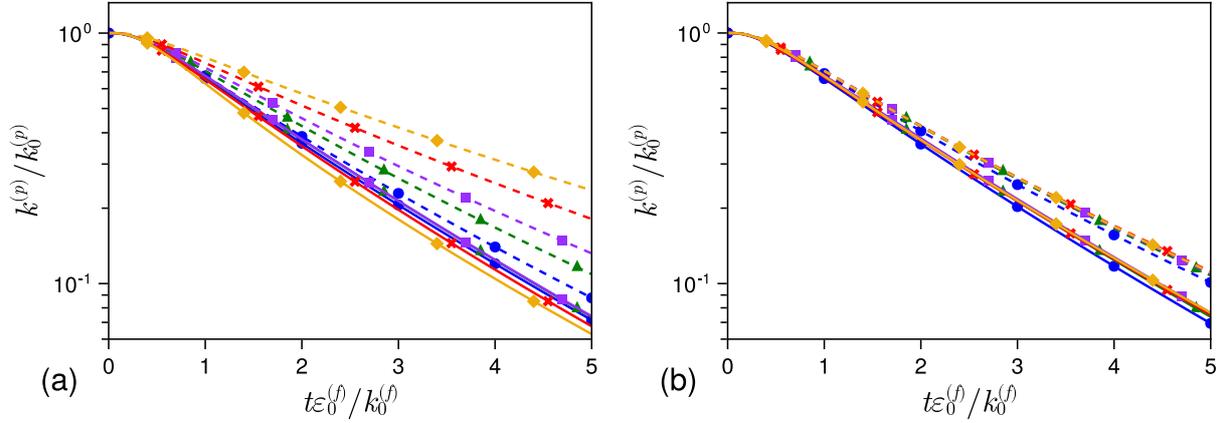}
    \caption{Temporal evolution of normalized particle kinetic energy decay in the $Re_{\lambda,0} = 52$, $St=100$ case using the (a) grid-based \roma interpolation kernel and (b) particle size--based \gaussone interpolation kernel. Solid lines represent cases with the Zonal-ADR model while dashed lines represent the uncorrected cases. Different colors/symbols represent different grid sizes: (\protect\bluecircle)~$D_p/\Delta=0.477$, (\protect\greentriangle)~$D_p/\Delta=0.716$, (\protect\purplesquare)~$D_p/\Delta=0.955$, (\protect\redx)~$D_p/\Delta=1.432$, (\protect\golddiamond)~$D_p/\Delta=1.91$.}
    \label{fig:Re52_KEp_St100}
\end{figure}

Similarly, figure \ref{fig:Re52_KEp_St100}(a,b) shows the temporal evolution of the particle kinetic energy for $St=100$ normalized by its initial value for both the Zonal-ADR corrected and uncorrected cases calculated using the (a) grid-based \roma and (b) particle size--based \gaussone interpolation kernels for different grid sizes. 
These results are consistent with the lower Reynolds number cases in Section~\ref{subsubsec:low_hit}. As explained in the prior section, in the absence of any self-disturbance correction, decreasing grid size relative to particle size in conjunction with a compact, grid-size-based interpolation kernel results in the underprediction of forces due to the highly localized sampling of largely disturbed fluid velocity. As the grid is refined, the interpolation kernel becomes even smaller relative to the particle, hence this underprediction of the force is magnified with each grid refinement, resulting in particle, and to a lesser extent fluid, statistics that diverge with grid refinement. Adding the Zonal-ADR correction to the grid-based interpolation kernel model leads to grid-converged predictions of particle kinetic energy for all grid resolutions. Small variations can still be seen in the corrected results for some grid sizes, but these differences are small, especially considering they are seen on a semi-log scale. Comparing it to the particle size--based \gaussone filter, grid-converged statistics for both the Zonal-ADR corrected and the uncorrected cases are observed. However, even though the uncorrected cases show grid convergence, the converged value differs from the converged value obtained with the Zonal-ADR correction scheme. Without any correction, there is still a slight underprediction of the force on the particles based on the undisturbed fluid velocity which results in this difference. Such a difference is small and only appears at later times.

\begin{figure}
    \centering
    \includegraphics[width=1.0\textwidth]{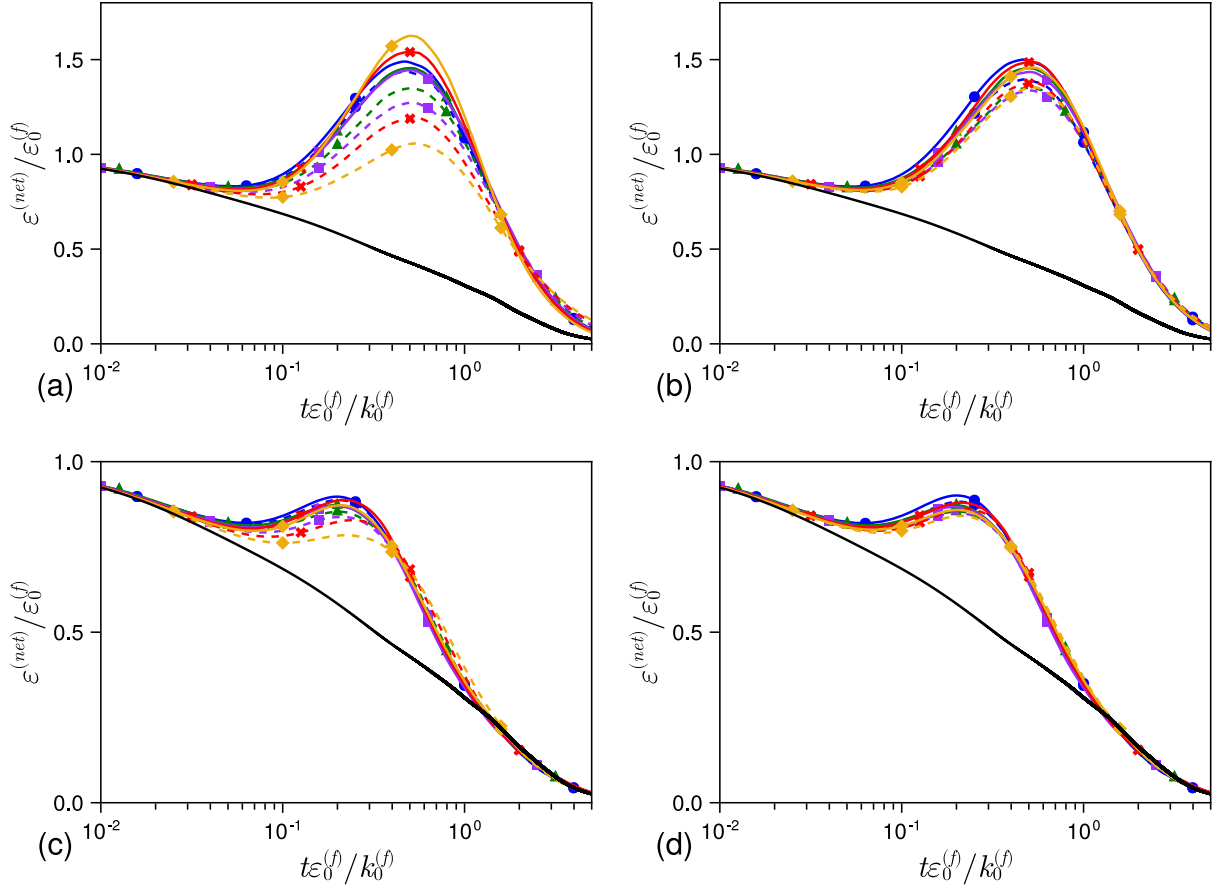}
    \caption{Temporal evolution of net normalized dissipation rate in the $Re_{\lambda,0} = 52$ for $St=100$ (top panel) and $St=10$ (bottom panel): (a,c) grid-based \roma interpolation kernel and (b,d) particle size--based \gaussone interpolation kernel. Solid lines represent cases with the Zonal-ADR model while dashed lines represent the uncorrected cases. Different colors/symbols represent different grid sizes: (\protect\bluecircle)~$D_p/\Delta=0.477$, (\protect\greentriangle)~$D_p/\Delta=0.716$, (\protect\purplesquare)~$D_p/\Delta=0.955$, (\protect\redx)~$D_p/\Delta=1.432$, (\protect\golddiamond)~$D_p/\Delta=1.91$. The normalized fluid dissipation rate of the unladen case is included as a reference~(\protect\blackline{}).}
    \label{fig:Re52_dkmixdt}
\end{figure}

The effect of the Zonal-ADR correction and choice of interpolation kernel across grid refinement on the net dissipation rate is examined next. As previously discussed, due to the nature of point-particle modeling, the net dissipation is not fully captured as resolved fluid dissipation. Instead, as with the lower Reynolds number cases, the net dissipation rate is calculated as the rate of change of the mixture kinetic energy (left-hand side of Eq. [\ref{eq:dissipation}]) \citep{mehrabadi2018direct, horwitz2020}. Figure \ref{fig:Re52_dkmixdt}(a,b) shows the net dissipation rate in the $St=100$ cases normalized by its initial value for both Zonal-ADR corrected and uncorrected runs calculated using the (i) grid-based \roma and (ii) particle size--based \gaussone interpolation kernels for different grid sizes. Once again, results are diverging with grid refinement when using an uncorrected model with the grid-based \roma interpolation kernel. Here, increasing the resolution relative to the particle size results in a lower net dissipation rate in the uncorrected \roma cases, with significant underpredictions at the higher grid resolutions. Sticking with the \roma interpolation kernel, if the model employs the Zonal-ADR correction, results are more grid converged than the uncorrected cases, but now the two highest resolution cases seem to predict slightly larger peak net dissipation than the other cases. As shown in the stationary particle case, when the grid size is smaller than the particle size and a grid-based \roma kernel is used, the two-way--coupled velocity field at the particle location can be unphysical, even with a correction scheme. This may contribute to the larger peaks in the net dissipation rate. Switching to the particle size--based \gaussone interpolation kernel, however, with the Zonal-ADR correction, the two-way coupled velocity field is more physical as also shown in the stationary sphere case. This 
seems to eliminate this difference, and similar predictions are seen across all grid resolutions, indicating that the corrected grid-based \roma model for the two highest-resolution cases was overpredicting net dissipation.  Similar to the previous results, using the \gaussone kernel with no correction model also seems to produce grid converged predictions of net dissipation rate, however, they are consistently and slightly lower than the predicted net dissipation when using the Zonal-ADR correction. Figure~\ref{fig:Re52_dkmixdt}(c,d) shows the same plots for the $St=10$ case. The main patterns remain consistent, but the differences are much smaller due to the low mass loading in these cases. One feature that is not present in this lower Stokes case is the deviation of the two highest resolution cases in the Zonal-ADR corrected, \roma interpolation kernel case (Figure~\ref{fig:Re52_dkmixdt}[c]). The poorer performance of the \roma kernel when particles are larger than the grid may have been mitigated due to the smaller effect of particles in general for this case.

\begin{figure}
    \centering
    \includegraphics[width=1.0\textwidth]{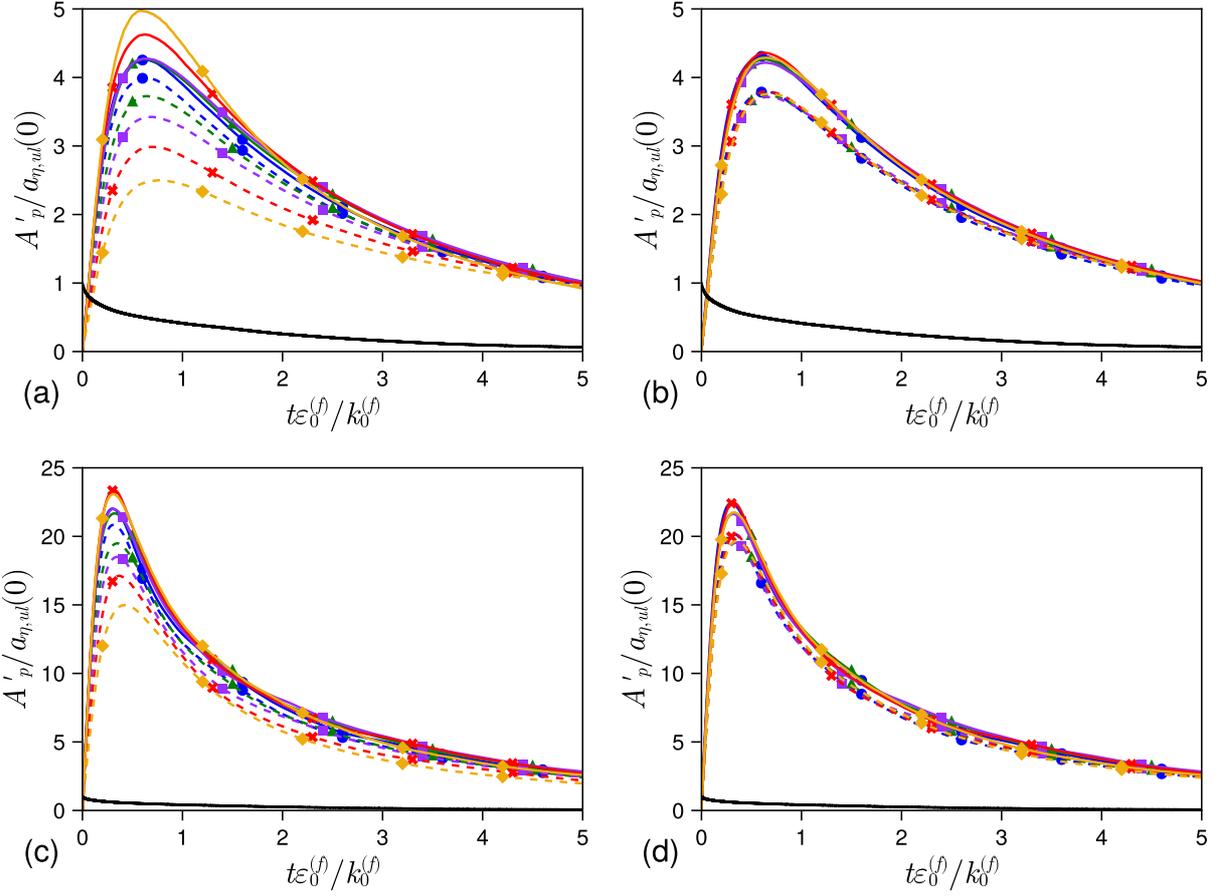}
    \caption{Temporal evolution of normalized particle rms acceleration in the $Re_{\lambda,0} = 52$ for $St=100$ (top panel) and $St=10$ (bottom panel): (a,c) grid-based \roma interpolation kernel and (b,d) particle size--based \gaussone interpolation kernel. Solid lines represent cases with the Zonal-ADR model while dashed lines represent the uncorrected cases. Particle rms acceleration is normalized by the initial Kolmogorov acceleration of the corresponding unladen case. Different colors/symbols represent different grid sizes: (\protect\bluecircle)~$D_p/\Delta=0.477$, (\protect\greentriangle)~$D_p/\Delta=0.716$, (\protect\purplesquare)~$D_p/\Delta=0.955$, (\protect\redx)~$D_p/\Delta=1.432$, (\protect\golddiamond)~$D_p/\Delta=1.91$. The normalized Kolmogorov acceleration of the unladen case is included as a reference~(\protect\blackline{}).}
    \label{fig:Re52_ap_rms}
\end{figure}

As previously seen, the effect of correcting for particle self-disturbance fields, especially at the volume loading studied here, is more clearly evident in particle statistics than it is in fluid statistics alone. To that end, another important particle statistic that can be examined is particle acceleration, as that is directly related to the particle force which is dependent on the force closures and hence the undisturbed fluid velocity in the relative slip-velocity between that particle and its surrounding fluid. Figure \ref{fig:Re52_ap_rms}(a,b) shows the particle Root mean square (rms) acceleration in the $St=100$ cases normalized by the initial Kolmogorov acceleration of the corresponding unladen case for both Zonal-ADR corrected and uncorrected runs calculated using the (i) grid-based \roma and (ii) particle size--based \gaussone interpolation kernels for different grid sizes. These particle acceleration statistics really highlight the effect of correction and interpolation kernel choice, potentially because acceleration involves the rate of change of velocity, whereas kinetic energy is more an integral quantity. A clear divergence in particle rms acceleration with decreasing grid size is seen in the uncorrected, grid-based \roma interpolation kernel results. 
Similar to the net dissipation results, the two highest resolutions actually end up overpredicting particle acceleration in the Zonal-ADR, grid-based \roma case, while the other three resolutions produce near-identical results. Notably, these two high resolutions which still deviate after using the Zonal-ADR correction with the \roma kernel are the two cases in which the particles are strictly larger than the grid ($D_p/\Delta = 1.432$ and $1.91$). As mentioned earlier, this again can be attributed to the overprediction of the force on the particle, just like in the stationary particle case with \roma kernel wherein the overprediction of the force results in negative two way--coupled velocity at the particle location.

The middle case ($D_p/\Delta = 0.955$) shows grid converged results with the other two coarser grids. This is potentially indicating that even with a correction scheme, using a compact, grid-based interpolation kernel can introduce errors when the particle is larger than the grid size. In contrast to this, when using the particle size--based \gaussone interpolation kernel, the particle rms acceleration statistics show excellent grid convergence across all grid sizes for both the uncorrected and the Zonal-ADR corrected cases. This highlights the importance of a particle size--based interpolation kernel when particles are on the order of the grid size, and especially when larger than the grid, even after correcting for the particle self-disturbances. The impact of the Zonal-ADR correction over the uncorrected approach is still highlighted very clearly here, with the uncorrected results showing consistent underprediction in particle rms acceleration relative to the Zonal-ADR corrected results. Figure~\ref{fig:Re52_ap_rms}(c,d) show the same particle rms acceleration plots for the $St=10$ cases. Here, the general patterns observed in the $St=100$ case hold true. While the mass loading is much lower, the effect of correction and interpolation kernel choice are still highlighted very well in particle acceleration. The main difference here is that the lower Stokes number is resulting in much larger accelerations than in the $St=100$ case, but this shows that the results observed are consistent across both Stokes numbers.

\begin{figure}[!htpb!]
    \centering
    \includegraphics[width=1.0\textwidth]{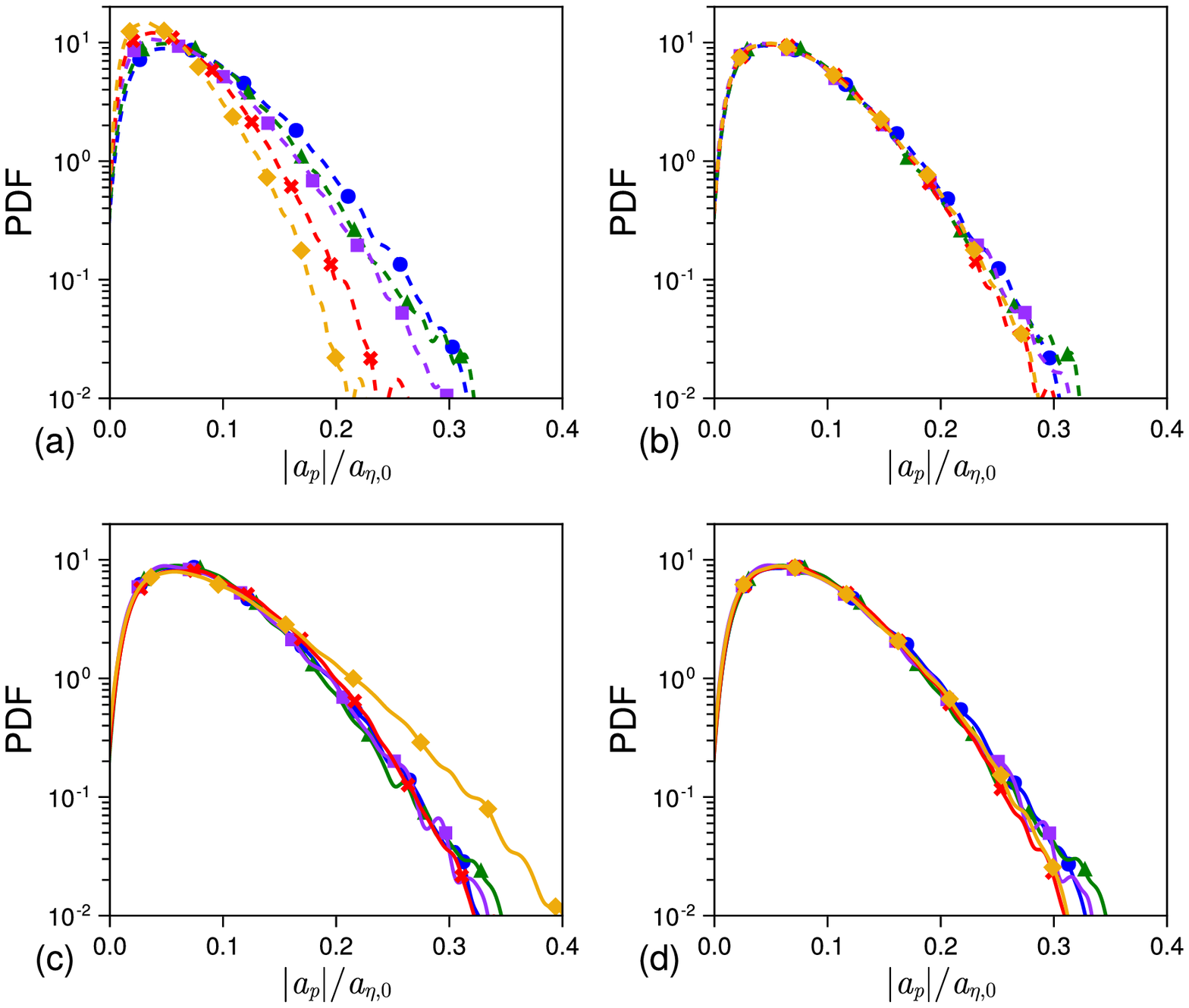}
    \caption{Particle acceleration PDFs in the $Re_{\lambda,0} = 52$, $St=100$ case at normalized time $t\epsilon_0^{(f)}/k_0^{(f)} \approx 0.92$ using the (a) uncorrected model with \roma interpolation kernel, (b) uncorrected model with \gaussone interpolation kernel, (c) Zonal-ADR model with \roma interpolation kernel, and (d) Zonal-ADR model with \gaussone interpolation kernel. Particle acceleration is the magnitude using all three components and is normalized by the initial Kolmogorov acceleration of the corresponding unladen case. Different colors/symbols represent different grid sizes: (\protect\bluecircle)~$D_p/\Delta=0.477$, (\protect\greentriangle)~$D_p/\Delta=0.716$, (\protect\purplesquare)~$D_p/\Delta=0.955$, (\protect\redx)~$D_p/\Delta=1.432$, (\protect\golddiamond)~$D_p/\Delta=1.91$.}
    \label{fig:Re52_ap_PDF_St100}
\end{figure}

Particle acceleration statistics are investigated further by examining the distribution of individual particle acceleration events. Figure~\ref{fig:Re52_ap_PDF_St100}(a-d) shows the probability density functions (PDFs) of particle acceleration magnitude for the $Re_{\lambda, 0} \approx 52$, $St=100$ case for the (a) uncorrected model with \roma interpolation kernel, (b) uncorrected model with \gaussone interpolation kernel, (c) Zonal-ADR corrected model with \roma interpolation kernel, and (d) Zonal-ADR corrected model with \gaussone interpolation kernel. When the grid-based \roma interpolation kernel is used in the absence of any correction, the PDF becomes narrower with increased grid resolution. Thus, when particles increase in size relative to the grid size, there are fewer high acceleration events and, as verified in particle rms acceleration, the peak acceleration events are smaller. Using the particle size--based \gaussone interpolation kernel with the uncorrected model results in grid-converged PDFs. In the Zonal-ADR case with \roma interpolation kernel, the correction results in grid converged PDFs for all cases except the most resolved case. Here, this case results in more high acceleration events than it should, which was also seen in the larger peak particle rms acceleration for the same case. Combining the Zonal-ADR correction with the particle size--based \gaussone interpolation kernel results in grid converged results with a slightly wider tail than the converged PDFs of the uncorrected \gaussone case, again indicating the importance of the Zonal-ADR correction scheme in addition to the particle size--based interpolation kernel.



\section{Conclusions}
\label{sec:conclusions}
In several particle-laden flow computations with direct or large-eddy simulation of the fluid flow, particle sizes comparable to the grid size or the smallest resolved flow scale are commonly encountered, e.g., sprays and droplets near the injector nozzle~\citep{moin2006large,apte2009large}, wall-bounded particle-laden flows~\citep{ferrante2005}, or transport of suspended sediments~\citep{finn2016}. Use of the point-particle approach under such situations, although not originally developed to be used in this limit, is common as it allows computation of large numbers of particle trajectories. 
However, application of a point-particle approach under the situation of the particle size being comparable to Kolmogorov scales needs additional considerations. These considerations include correcting for the self-disturbance field and carefully choosing the interpolation kernels for interaction forces in the two way--coupling. 
This work thoroughly evaluates the effect of the aforementioned considerations on the accuracy of the point-particle models in the limit of $D_p/\Delta \sim {\mathcal O}(1)$.

It is shown that the grid-based interpolation kernels, which vary based on the local grid resolution, irrespective of the particle size, significantly underpredict the interaction force and hence the decay rates of particle kinetic energy and magnitude of particle acceleration, especially when no model is used for the self-disturbance correction. Furthermore, as these grid-based kernels only depend on grid size, a clear divergence in fluid and particle statistics is observed with grid refinement in the absence of any particle self-disturbance correction. Particle size--based kernels, wherein the kernel widths scale with the particle size, can better capture these interactions even without any correction model, producing grid-converged results. As the grid is refined and the particle size becomes larger than the grid size, a kernel width proportional to particle size keeps the region of particle-fluid interaction unchanged and allows sampling of the fluid parameters from a region that is less affected by the self-disturbance field. 
While convergent under grid refinement, the particle kinetic energy decay rate, net dissipation rate, and magnitude of particle rms acceleration are slightly under-predicted with no self-disturbance correction model. With correction for self-disturbance, both the grid-based and the particle size--based kernels provides grid-convergent results. Therefore, while the use of a particle size--based interpolation kernel may mitigate some of the errors resulting from the use of the two way--coupled disturbed fluid velocity in the force closures, there is still a need for self-disturbance correction for particles comparable to the grid size. However, when particles become larger than the grid size, the grid-based kernels distribute the force computed based on the undisturbed fluid velocity in a narrow region, resulting in a locally large value of the force, even though the global value is conserved. This increased force can result in an unphysical two-way--coupled velocity at the particle location with the grid--based kernel. This is also clearly observed in the stationary particle case with grid-based \roma kernel for particles larger than the grid and with correction for the self-disturbance. Such an effect was shown to be absent with the particle--based kernel, even for particles larger than the grid size.

These results suggest that the use of a particle size--based interpolation kernel in conjunction with a correction model for the self-disturbance field is recommended to obtain the best results. With particle size--based kernels, the predictions without self-disturbance corrections are grid convergent, but the turbulent kinetic energy and dissipation rate are slightly underpredicted. Implementing the particle size--based interpolation kernel requires knowledge of several neighboring control volumes, beyond the nearest neighbors surrounding the particle, especially if the particle size is comparable to the grid size. For complex unstructured grids and parallel computing, this can lead to increased memory and computational time. In such cases, grid-based kernels that depend only on nearest neighbors are a  practical option, provided a self-disturbance correction model is used to obtain consistent and grid-converged results. This is especially important for any application which involves particle-laden flow near boundaries, as computational grids are refined in the wall-normal direction to resolve the flow and thus particles can range from being subgrid to larger than the grid size within the same domain.

\subsection*{Acknowledgments}
Computing time on TACC's Frontera and Purdue's Anvil is appreciated. N.~K. and S.~V.~A. acknowledge NSF award\#1851389. S.~S.~J. acknowledges funding support from Boeing Co. Authors thank Jeremy Horwitz, Mohammad Mehrabadi, and Shankar Subramaniam for the PR-DNS data. A preliminary version of this work has been published \citep{Apte2022} as the Center for Turbulence Research Proceedings of the Summer Program (CTR-PSP) and is available online\footnote{https://web.stanford.edu/group/ctr/ctrsp22/iii02\_Apte.pdf}.


\bibliography{manuscript}

\end{document}